\documentclass[structabstract]{aa}

\usepackage{graphicx}
\usepackage{txfonts}
\usepackage{natbib}
\bibpunct{(}{)}{;}{a}{}{,}

\newcommand{\feh}{\mathrm{[Fe/H]}}
\newcommand{\teff}{T_\mathrm{eff}}
\newcommand{\logg}{\log g}

\newcommand{\fei}{Fe~\textsc{i}}
\newcommand{\feii}{Fe~\textsc{ii}}

\newcommand{\asset}{Ass$\epsilon$t}
\newcommand{\ms}{m~s$^{-1}$}
\newcommand{\vsini}{V\sin i}

\begin{document}

\defcitealias{kdwarfs-p1}{Paper~I}
\defcitealias{kdwarfs-p2}{Paper~II}
\defcitealias{kdwarfs-p3}{Paper~III}

\title{Granulation in K-type dwarf stars}
\subtitle{II. Hydrodynamic simulations and 3D spectrum synthesis}
\titlerunning{Granulation in K-dwarfs. II.}
\authorrunning{Ram\'{\i}rez et~al.}

\author{I. Ram\'{\i}rez   \inst{1,2} \and
        C. Allende Prieto \inst{2,3} \and
        L. Koesterke       \inst{4}  \and
        D. L. Lambert      \inst{2}  \and
        M. Asplund         \inst{1}
       }

\institute{Max Planck Institute for Astrophysics,
           Postfach 1317, 85741 Garching, Germany
           \and
	   McDonald Observatory and Department of Astronomy,
           University of Texas, Austin, TX 78712-0259, USA
           \and
	   Mullard Space Science Laboratory, University College London,
	   Holmbury St.~Mary, Dorking, Surrey RH5 6NT, UK
           \and
           Texas Advanced Computing Center, 
           University of Texas, J.~J.~Pickle Research Campus, 
           Austin, TX 78758-4497, USA \\
           \email{ivan@mpa-garching.mpg.de}
          }

\date{Received January 29, 2009; accepted May 7, 2009}

\abstract
{}
{To explore the impact of surface inhomogeneities on stellar spectra, granulation models need to be computed. Ideally, the most fundamental characteristics of these models should be carefully tested before applying them to the study of more practical matters, such as the derivation of photospheric abundances. Our goal is to analyze the particular case of a K-dwarf.}
{We construct a three-dimensional radiative-hydrodynamic model atmosphere of parameters $\teff=4820$\,K, $\logg=4.5$, and solar chemical composition. Using this model and 3D spectrum synthesis, we computed a number of \fei\ and \feii\ line profiles. The observations presented in the first paper of this series were used to test the model predictions. The effects of stellar rotation and instrumental imperfections are carefully taken into account in the synthesis of spectral lines.}
{The theoretical line profiles show the typical signatures of granulation: the lines are asymmetric, with their bisectors having a characteristic C-shape and their core wavelengths shifted with respect to their laboratory values. The line bisectors span from about 10 to 250~\ms, depending on line strength, with the stronger features showing larger span. The corresponding core wavelength shifts range from about $-200$~\ms\ for the weak \fei\ lines to almost $+100$~\ms\ in the strong \fei\ features. Based on observational results for the Sun, we argue that there should be no core wavelength shift for \fei\ lines of $EW\gtrsim100$~m\AA. The cores of the strongest lines show contributions from the uncertain top layers of the model, where non-LTE effects and the presence of the chromosphere, which are important in real stars, are not accounted for. The \feii\ lines suffer from stronger granulation effects due to their deeper formation depth which makes them experience stronger temperature and velocity contrasts. For example, the core wavelength shifts of the weakest \feii\ lines are about $-600$~\ms. The comparison of model predictions to observed \fei\ line bisectors and core wavelength shifts for our reference star, HIP~86400, shows excellent agreement, with the exception of the core wavelength shifts of the strongest features, for which we suspect inaccurate theoretical values. Since this limitation does not affect the predicted line equivalent widths significantly, we consider our 3D model validated for photospheric abundance work.}
{}


\maketitle

\section{Introduction}

Solar granulation is the visible manifestation of surface convection \citep[e.g.,][]{bray84,muller99}. When rising gas from the convective zone reaches the photosphere, it cools down by radiation losses. Since the opacity decreases steeply with decreasing gas temperature, the temperature of the photosphere drops abruptly as hydrogen recombines and the gas becomes suddenly transparent. The upflowing material in the granules decelerates as it enters the convectively stable region due to the higher pressures above and flows sideways until encountering other granules before being accelerated downwards due to it now being overdense and having negative buoyancy. Since convective envelopes are present in other cool stars (spectral type F and later on the main sequence, G and later for giants and supergiants), it is expected that they also experience surface granulation.

The physics of radiatively driven convection is complex; it is three-dimensional, time-dependent, non-local, and non-linear. Parameterized models of granulation may require a large number of free parameters whose physical interpretation is obscure. Furthermore, they do not guarantee a unique solution. On the other hand, the continuous growth of computer power is allowing us to solve this problem numerically using only the basic laws of hydrodynamics, including radiative transfer in the energy equation. A better understanding of solar and stellar granulation is thus possible through numerical simulations \citep[e.g.,][]{nordlund90,freytag96,stein98,asplund99,ludwig99,robinson03,voegler04,asplund05:review,collet06,collet07,trampedach07,nordlund08}.

Theoretical models can be computed at will, but using them is meaningless until we are reassured that they include the relevant physics behind the phenomenon they intend to represent. The validation of theoretical models, i.e., verifying that their most fundamental predictions are supported by targeted observations, is necessary before exploring the impact of their effects on other matters, such as the inferred chemical compositions in the case of stellar atmospheres.

Stellar disks are usually unresolved. The exceptions are nearby red supergiants, for which interferometric observations with present-day technology are in principle capable of revealing surface structures related to convection \citep[e.g.,][]{gilliland96,chiavassa08}.  In any case, there are no near-future prospects for direct observations of stellar granulation.

Fortunately, granulation signatures are also present in stellar spectra: net line wavelength shifts and asymmetries. Very careful observations are required to detect the effects of granulation in this manner (\citealt{dravins87:observability}; \citealt{kdwarfs-p1}, hereafter \citetalias{kdwarfs-p1}). High quality data have been used by several authors to show that stars with convective envelopes experience granulation with different degrees of intensity and velocity contrasts \citep[e.g.,][]{gray82,gray05,dravins87:line_asymmetries,dravins08,allende02}. Other velocity fields may become prominent in more evolved stars \citep{gray08}. Indeed, \cite{gray89} have defined a granulation boundary on the HR diagram based on observations of spectral line asymmetries in hot and cool stars. They find that the shape and magnitude of the line asymmetries in hot stars (A-type and hotter on the main sequence, F-type and hotter for the supergiants) are qualitatively different from those that correspond to solar granulation and argue that they cannot be related to deep envelope convection.

Spectrum synthesis using classical, hydrostatic model atmospheres predicts symmetric lines whose core wavelengths correspond exactly to the input laboratory wavelengths. In real stellar atmospheres that experience granulation, line profiles coming from bright, hot granules are blueshifted because the material is rising and therefore approaching the observer while line profiles coming from the dark, cool intergranular lanes are redshifted. The continuum level of the upflow line-profiles is higher than that of the line-profiles associated with the cool downflows. Thus, if the granulation pattern is unresolved, an observed line profile is dominated by the blueshifted component. The profile is then asymmetric and its observed core wavelength is blueshifted \citep[e.g.,][]{dravins81}. The line asymmetry can be quantified by the \textit{line bisector}, defined as the location of the midpoints of horizontal segments joining the blue and red wings of the lines, and the core wavelength shifts due to granulation are often referred to as \textit{convective blueshifts}.

The goal of this series of papers is to understand the surface inhomogeneities (granulation) present in K-dwarf stars and explore their impact on the determination of fundamental parameters and chemical compositions derived from the star's spectral energy distribution. In the first part of the series \citep[hereafter \citetalias{kdwarfs-p1}]{kdwarfs-p1}, we showed that, with very careful observing strategies, it is possible to not only detect but also accurately quantify the effects of granulation on line profiles in the spectra of a few bright K-type dwarf stars. A theoretical granulation model has been calculated and it is described in detail in this paper, along with the tests made to ensure that its most fundamental predictions, namely their impact on the shapes of absorption line profiles (specifically those due to \fei), are valid. This validation will allow us to explore with confidence the so-called ``3D effects'' on the determination of chemical compositions and fundamental parameters of K-dwarfs in the final part of this series \citep[hereafter \citetalias{kdwarfs-p3}]{kdwarfs-p3}.

\section{The 3D model} \label{s:3dmodel}

Assuming local thermodynamic equilibrium (LTE), we computed a three-dimensional radiative-hydrodynamic model atmosphere using the methods described in \cite{stein98}. A plane-parallel box of the stellar envelope was modeled by solving the fluid dynamics equations of mass continuity, momentum conservation, and energy conservation \cite[e.g., Eqs.~1--3 in][]{stein98}. The effect of temperature enters through the equation of state and by the radiative heating rate, which is determined by solving the radiative transfer equation \cite[e.g., Eq.~7 in ][]{stein98}.

In the calculation of the 3D model, the equation of state (as given by \citealt{mihalas88}), continuum opacities, and source functions were obtained by interpolation in the tables included in the updated MARCS stellar atmosphere package \citep{gustafsson75,gustafsson08}. The adopted line opacities (opacity distribution functions) are those by \cite{kurucz93:cd13,kurucz93:cd18}.

Surface gravity, chemical composition, and entropy of the gas entering the simulation box are the fundamental parameters of the 3D model calculation. The effective temperature is not an input parameter as in 1D models but it is rather adjusted by modifying the state of the gas entering the bottom of the computational domain until it reaches an average similar to the target value of 4820\,K. The other input parameters are $\logg=4.5$ and $\feh=0.0$. The solar abundances adopted are those given by \cite{grevesse98}.

The model consists of $150\times150\times82$ grid points representing a rectangular box of the following geometrical dimensions: $4.7\times4.7\times3.2$~Mm (1\,Mm~$\equiv10^6$\,m). The simulation box has periodic horizontal boundaries, gas is allowed to escape at the top, while the density and energy of the incoming gas at the bottom boundary are adjusted to conserve the entropy. After relaxation, the simulation was run for about one hour of stellar time, from which 100 snapshots separated by 40~s were extracted and employed for the line formation calculations presented in this study (the actual hydrodynamical time-step was approximately 0.2\,s).

\begin{figure*}
\centering
\includegraphics[width=3cm]{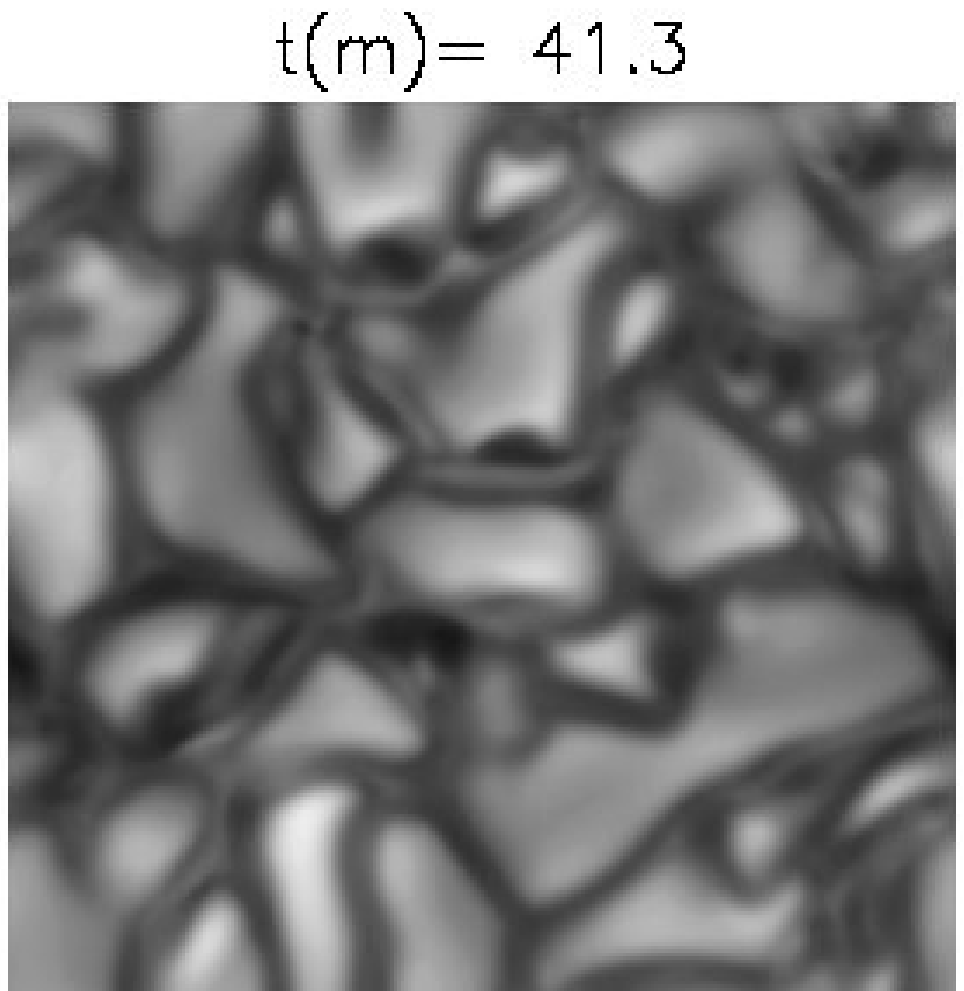}
\includegraphics[width=3cm]{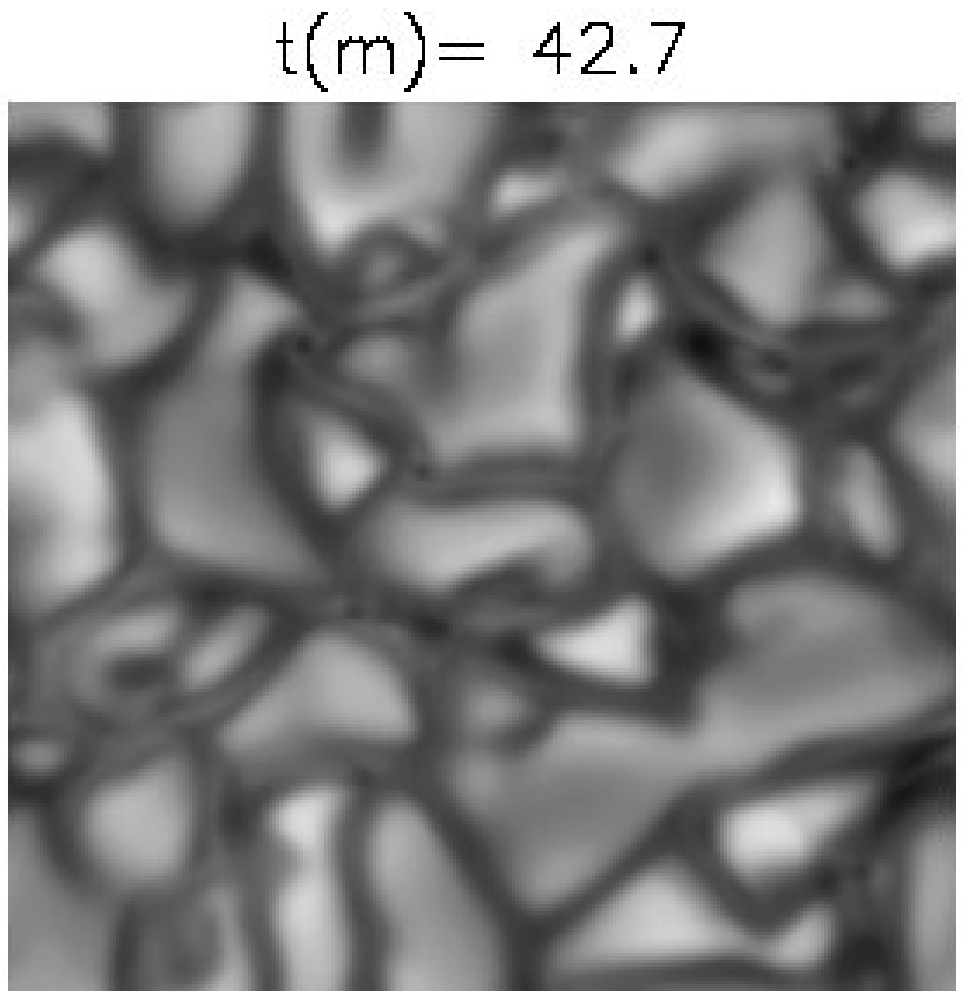}
\includegraphics[width=3cm]{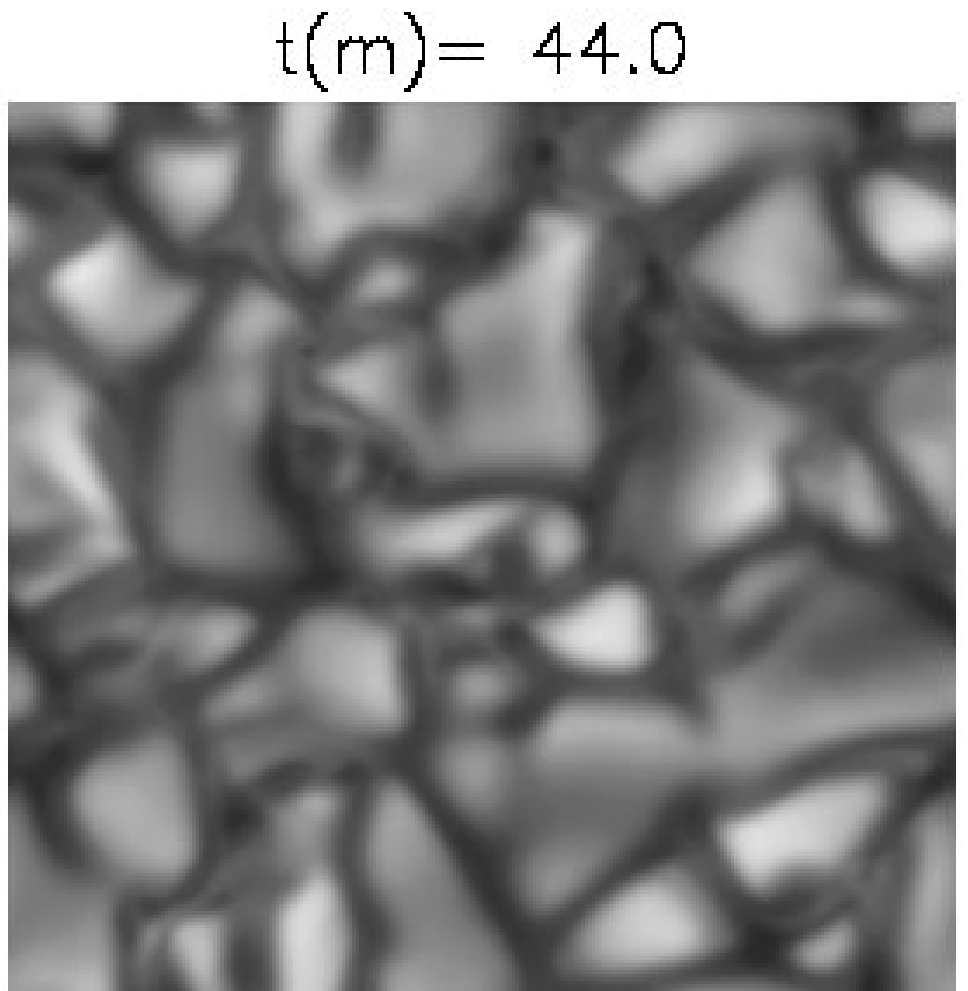}
\includegraphics[width=3cm]{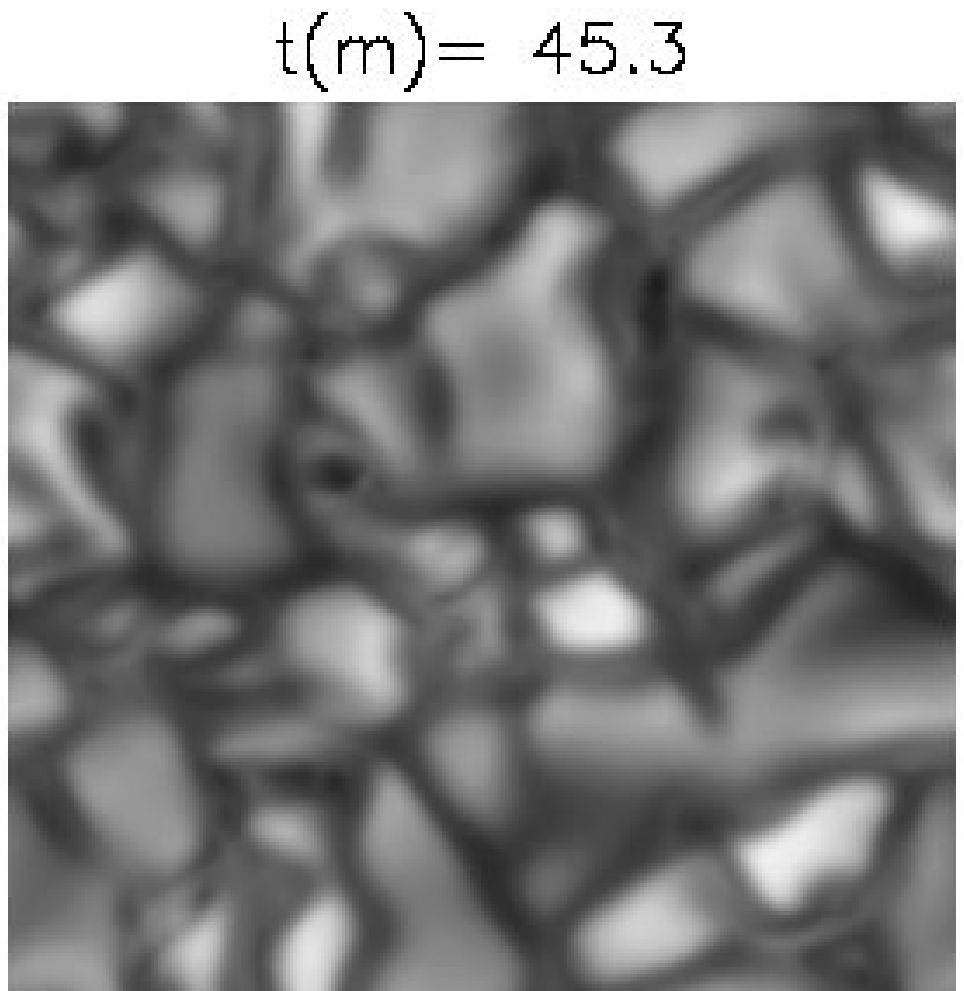}
\includegraphics[width=3cm]{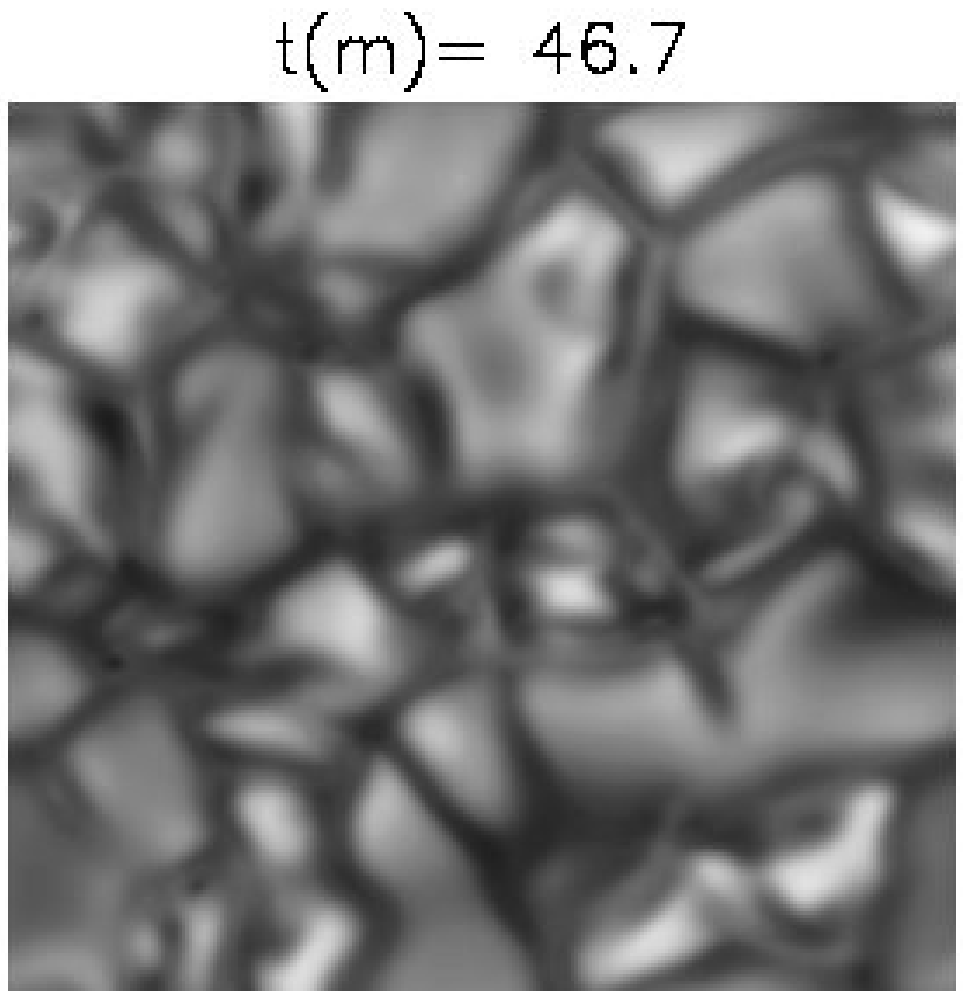}
\includegraphics[width=3cm]{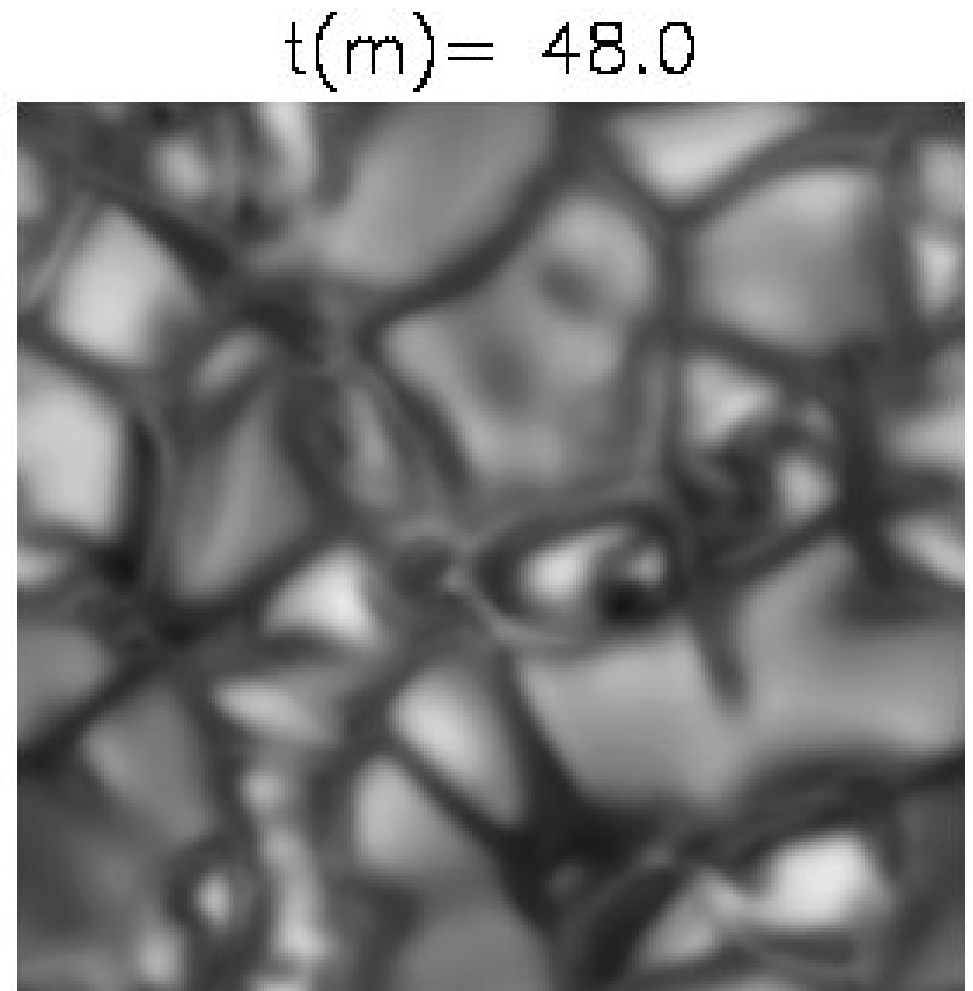}
\includegraphics[width=3cm]{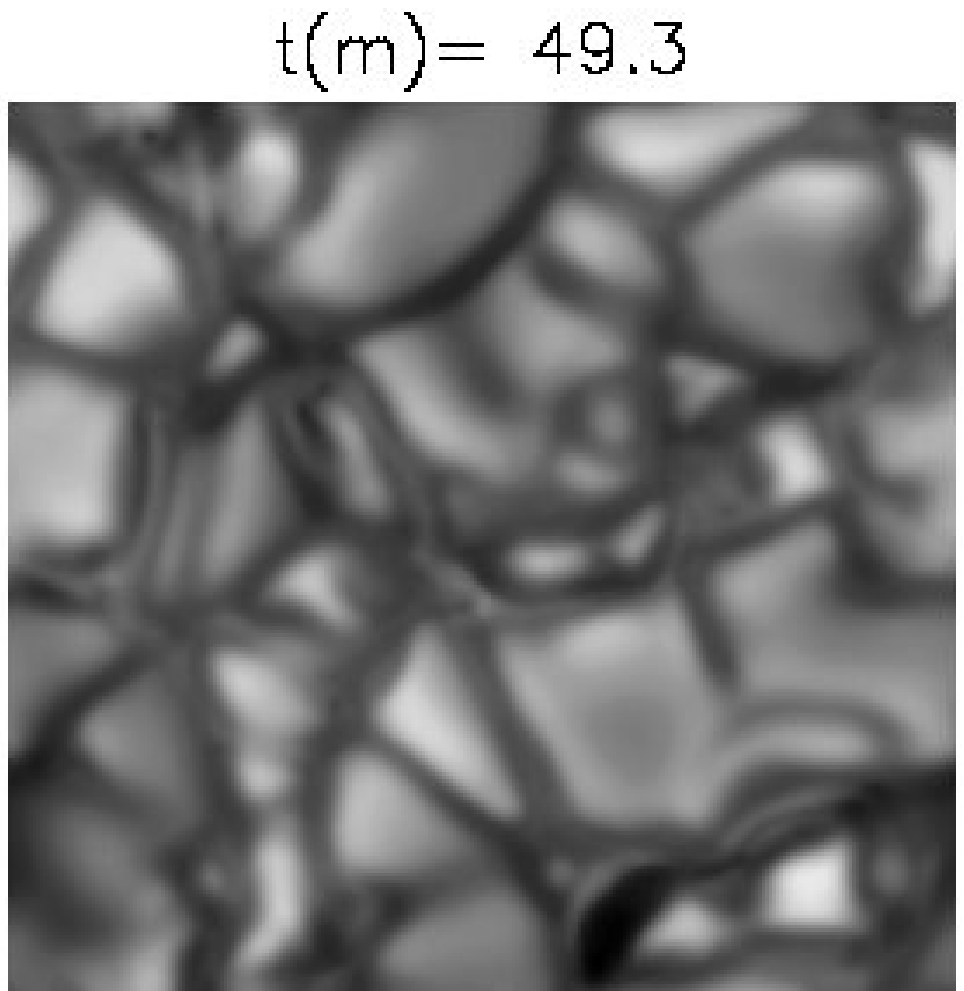}
\includegraphics[width=3cm]{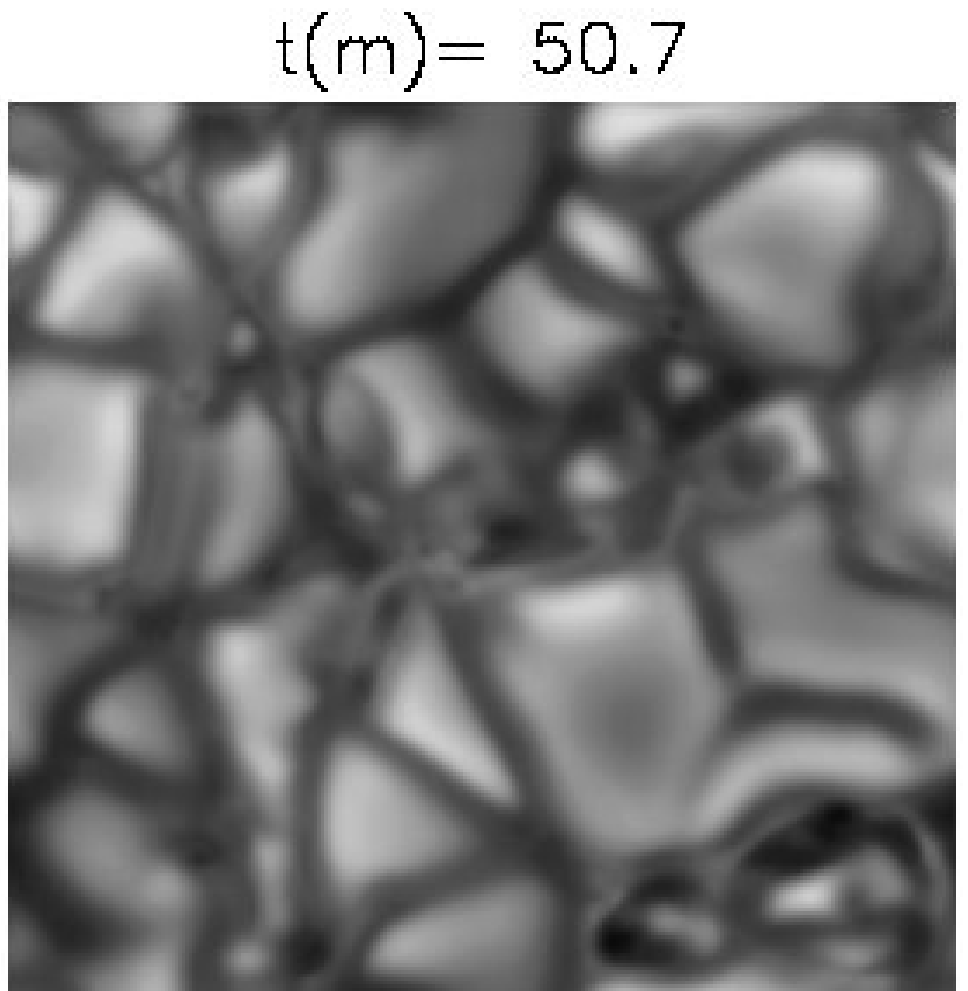}
\includegraphics[width=3cm]{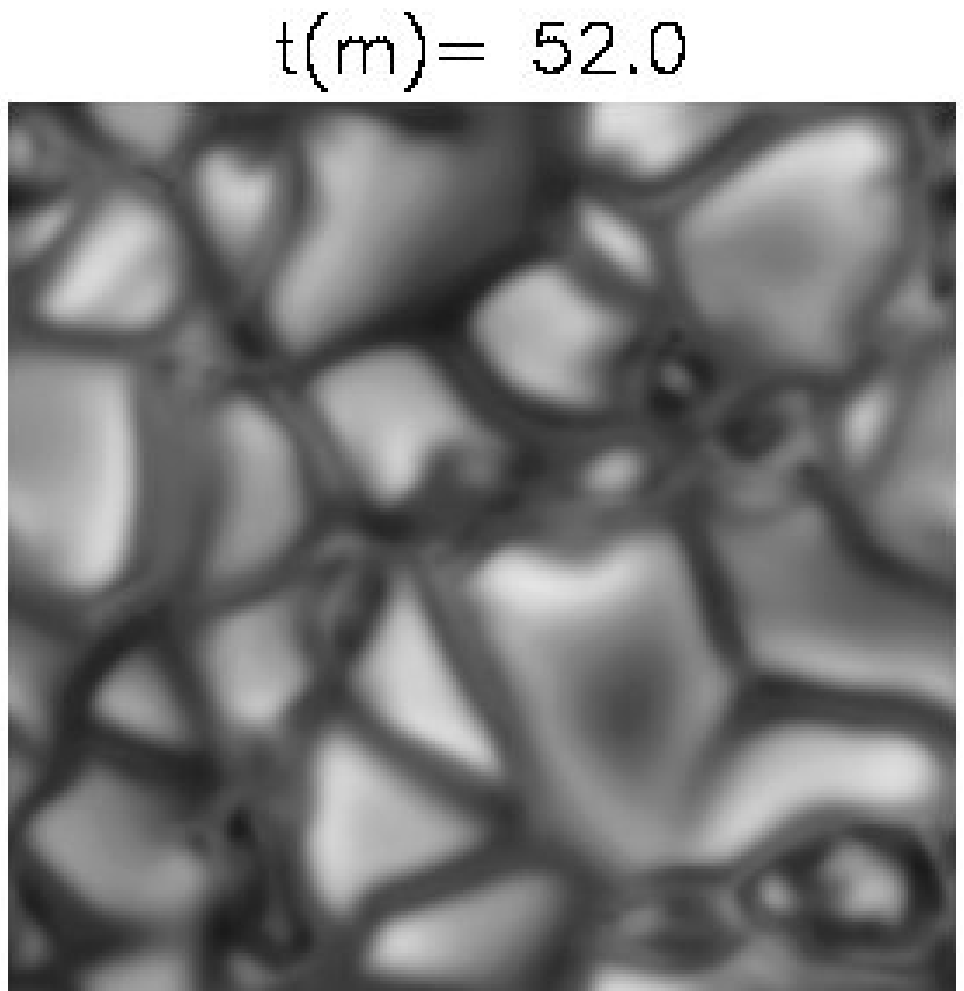}
\includegraphics[width=3cm]{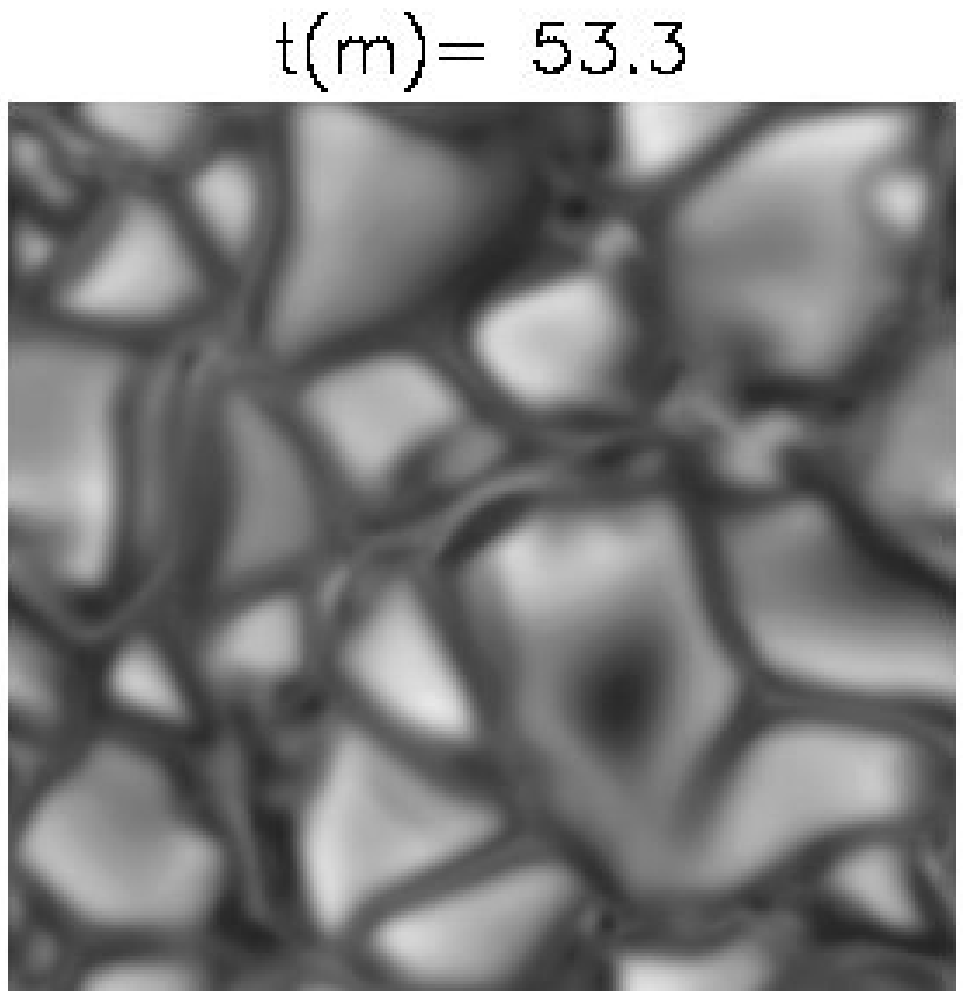}
\includegraphics[width=3cm]{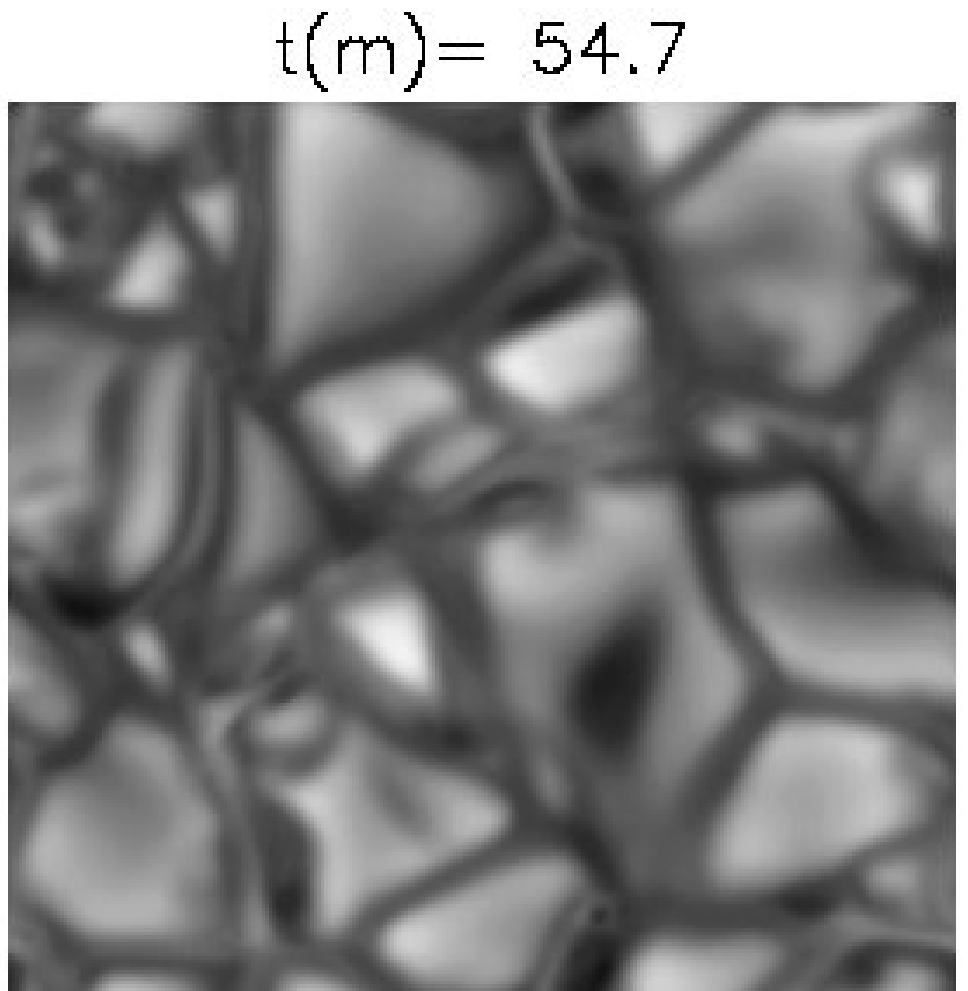}
\includegraphics[width=3cm]{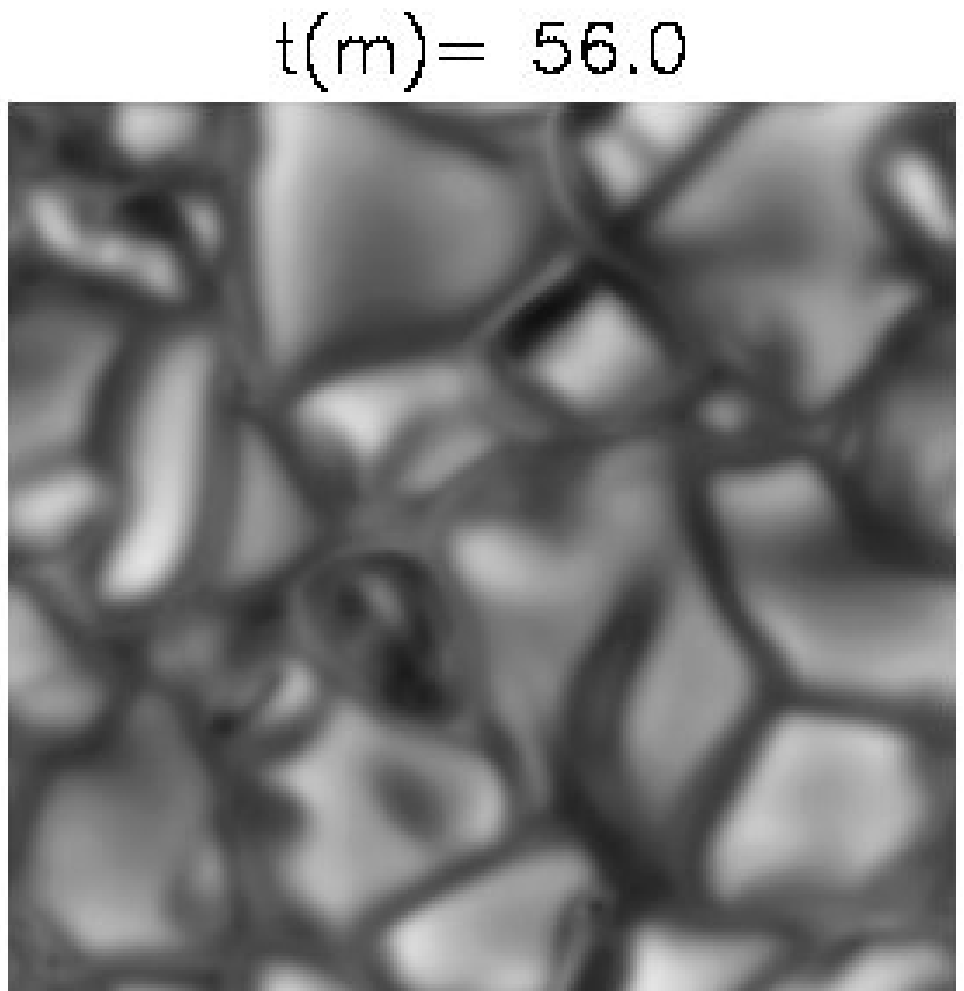}
\includegraphics[width=3cm]{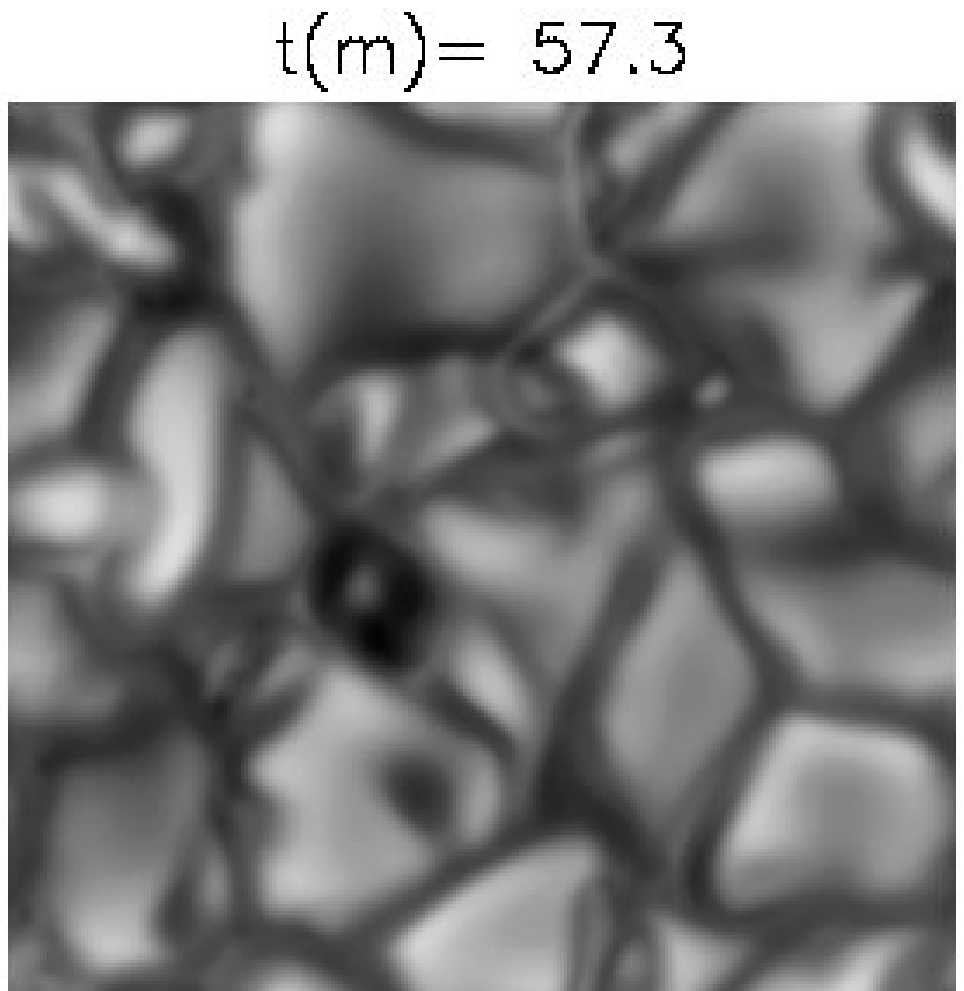}
\includegraphics[width=3cm]{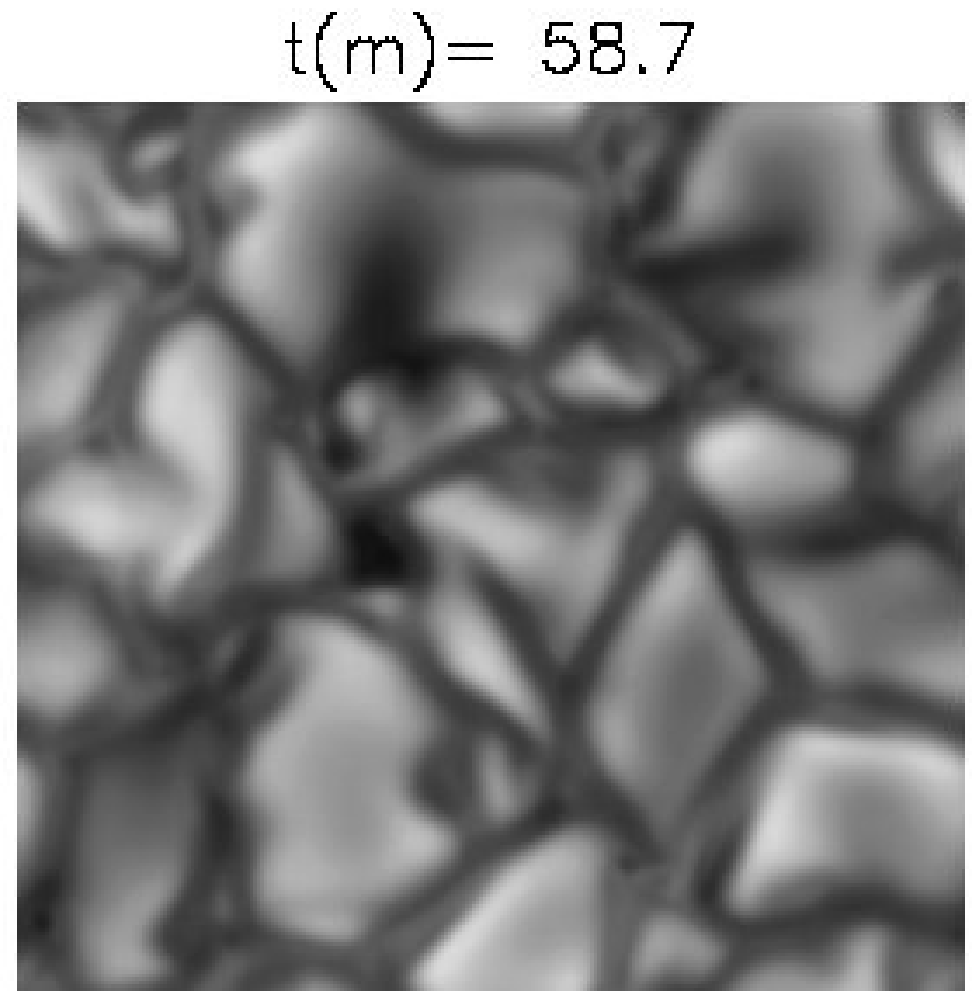}
\includegraphics[width=3cm]{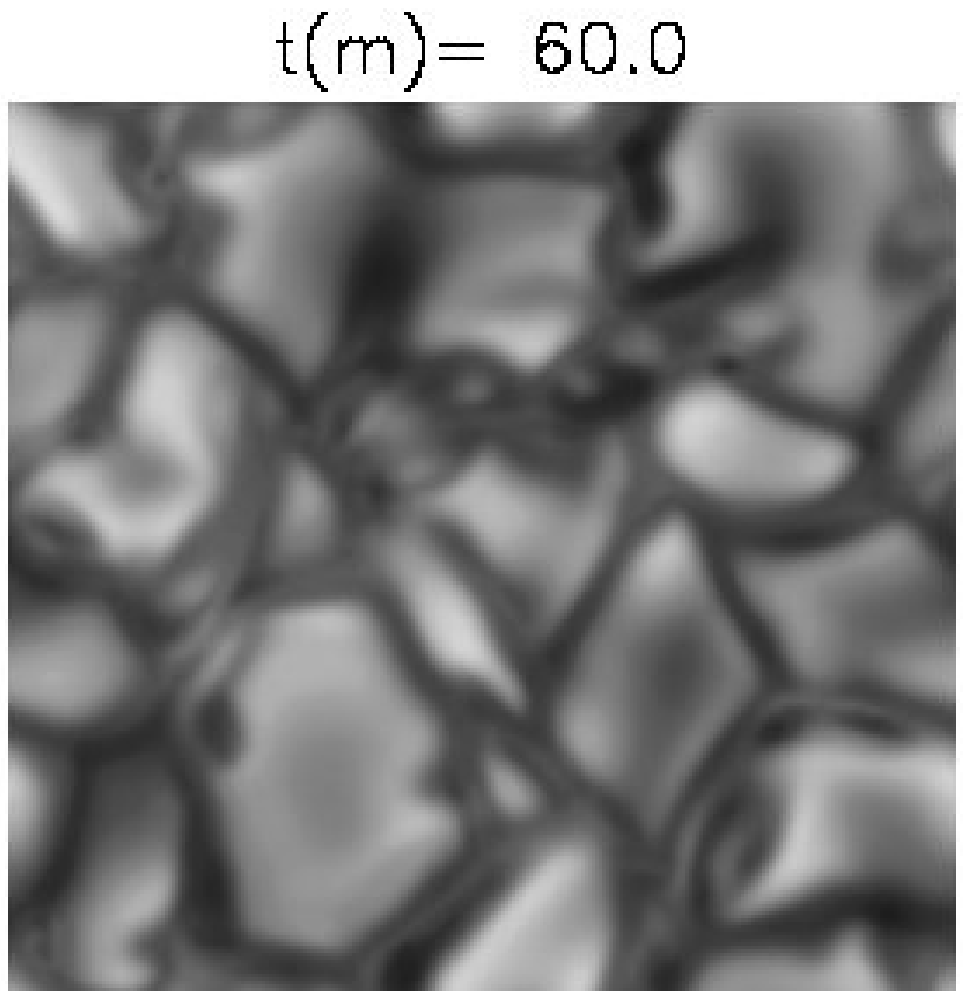}
\includegraphics[width=3cm]{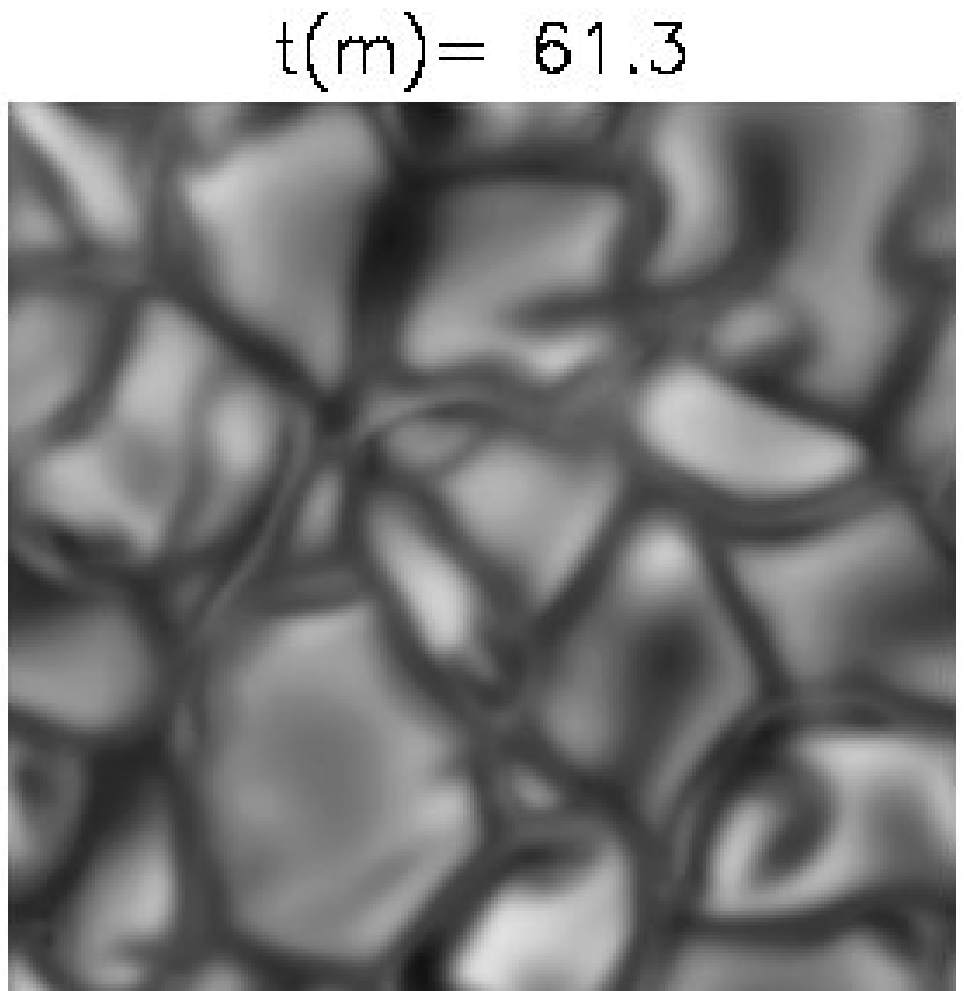}
\includegraphics[width=3cm]{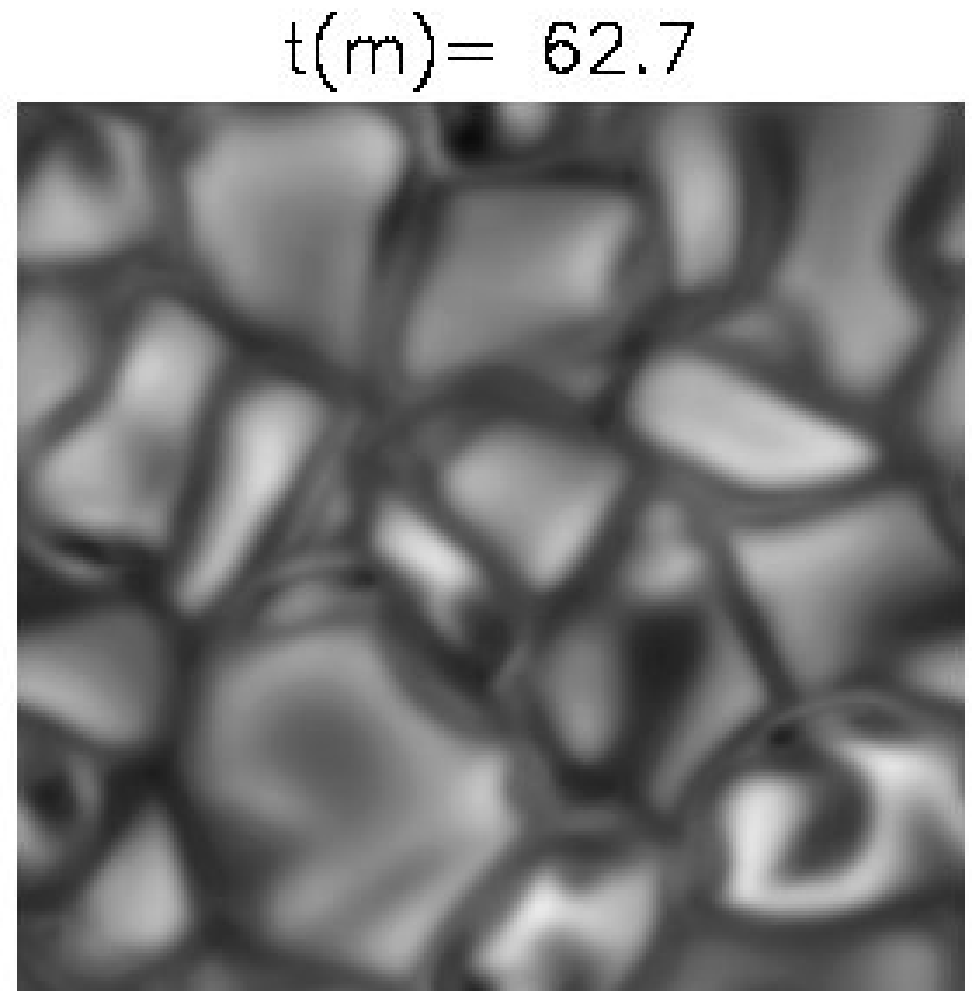}
\includegraphics[width=3cm]{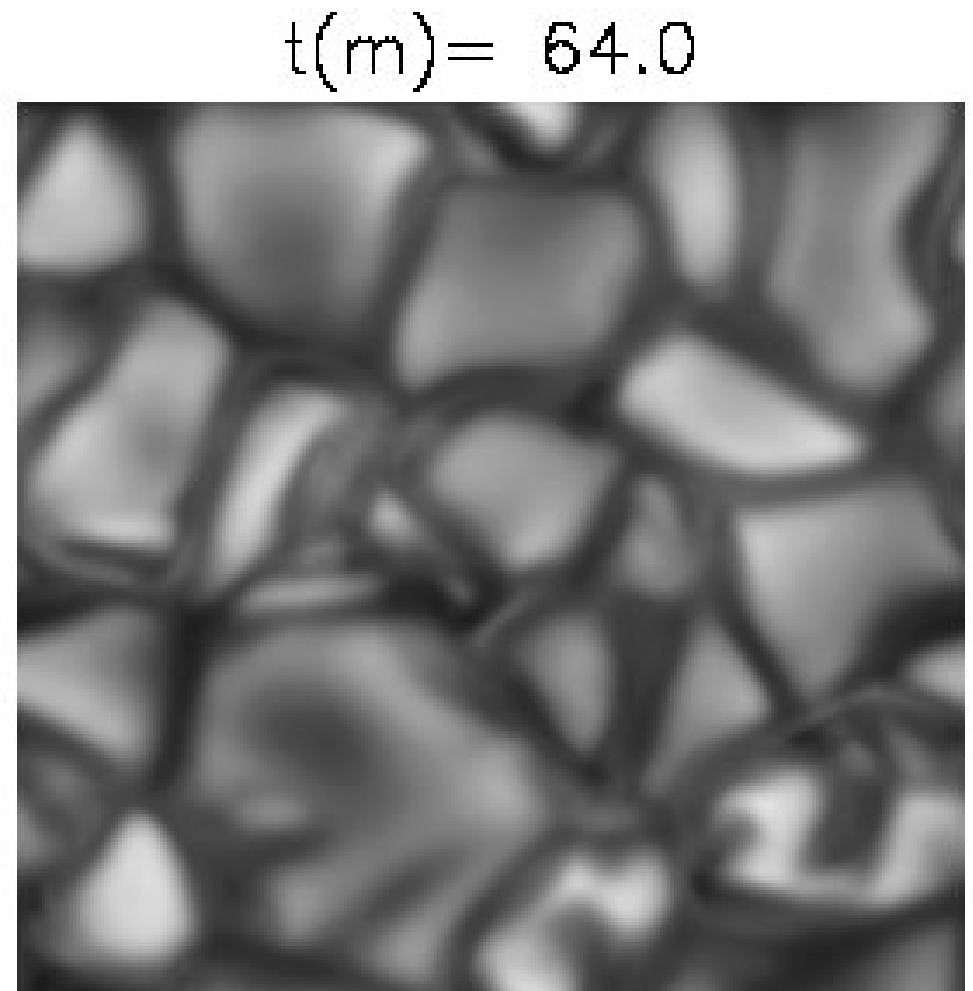}
\includegraphics[width=8cm,bb=54 590 337 643]{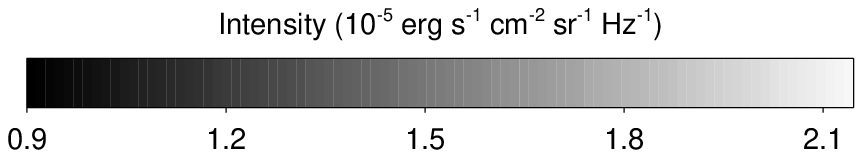}
\caption{Snapshots of the emergent intensity (continuum) at 6151~\AA\ predicted by our 3D K-dwarf model atmosphere (details on the spectrum synthesis are given in Sect.~\ref{s:3dsynthesis}). Each panel shows an area of $4.7\times4.7$\,Mm$^2$ which corresponds to the dimensions of a horizontal cross-section of the simulation box. The time, in minutes, is given on top of each panel. Only the last 22.7 minutes of the simulation are shown.}
\label{f:granulation_timeseries}
\end{figure*}

The time evolution of the emergent intensity that would be observed on the surface of this model star is illustrated in Fig.~\ref{f:granulation_timeseries} (in Sect.~\ref{s:3dsynthesis} we provide details on the calculation of the emergent intensity from the 3D model). The pattern of granules and intergranular lanes resembles that of the Sun but on a smaller geometrical scale. The correlation between the emergent intensity of a given snapshot and those of the subsequent ones decreases smoothly and vanishes for images separated by about 10 minutes, a number comparable to that obtained for the solar simulation. Thus, the lifetime of granules in our K-dwarf simulation is similar to that of the Sun.

\begin{figure}
\includegraphics[width=9.2cm,bb=70 370 558 700]{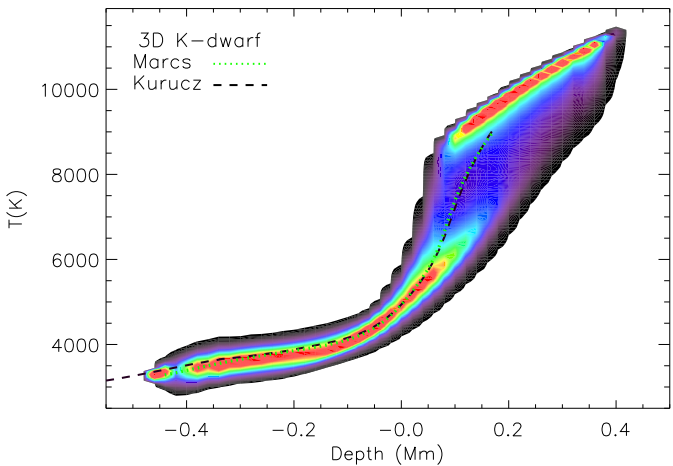}
\caption{Temperature structure of our 3D K-dwarf model atmosphere near the visible surface (the full simulation extends much deeper than shown here). The color code represents the number of grid points at a given temperature and depth, with red being the largest and dark blue the lowest number. The dotted green line and dashed black line correspond to temperature structures of 1D model atmospheres (see legend on the top left). The depth scale is set to zero at the geometrical plane where the mean Rosseland optical depth is equal to 1; i.e., at the ``visible'' surface.}
\label{f:temp3d}
\end{figure}

Fig.~\ref{f:temp3d} shows the temperature structure of our model in the upper 1~Mm layers (this is the portion that we use for the spectral line calculations, as explained in Sect.~\ref{s:3dsynthesis}). The rapid decrease of temperature with atmospheric height just below the visible surface ($\mathrm{depth}=0$),\footnote{The ``depth'' variable corresponds to the geometrical depth, increasing inward, and it is zero at the plane where the spatially-averaged Rosseland optical depth along the vertical direction is equal to one.} attributed to the feedback between radiation losses and decreasing opacity, is the most prominent feature. There, across a layer of only about 50\,km, the gas temperature drops about 3000~K. The apparent discontinuity of the temperature structure at $\mathrm{depth\simeq0.1}$\,Mm is due to this effect. In the regions of line formation ($\mathrm{depth}<0$), the temperature fluctuations have extremes that differ by about $1000$\,K.

Also shown in Fig.~\ref{f:temp3d} are one-dimensional models computed with the ATLAS \citep{kurucz79,kurucz93:cd13} and MARCS \citep{gustafsson75} codes.\footnote{The references given here correspond to those that describe the basic properties of the models. Several improvements have been made since their publication. The models presented in Fig.~\ref{f:temp3d} are those available at: {\tt http://kurucz.harvard.edu/grids.html} (the ``odfnew'' version) and {\tt http://marcs.astro.uu.se/} (see also \citealt{gustafsson08} for more details on the updated MARCS models).} It is clear that, in the regions of line-formation, 1D and 3D models show very similar temperature structures if, for the latter, the average value of the temperature is obtained at each depth. We must warn, however, that it is because of the temperature (and velocity) fluctuations that the 3D model is more realistic and that this ``similarity'' with 1D models should not be used to argue that 1D and 3D models of K-dwarf photospheres are equivalent. The reason why the average 3D and 1D temperature structures are similar is that 3D models are nearly in radiative equilibrium at solar metallicity \citep[e.g.,][]{stein98}, which is not the case for metal-poor stars \citep[e.g.,][]{asplund99}.

\begin{figure}
\includegraphics[width=6.9cm,bb=0 -10 600 610]{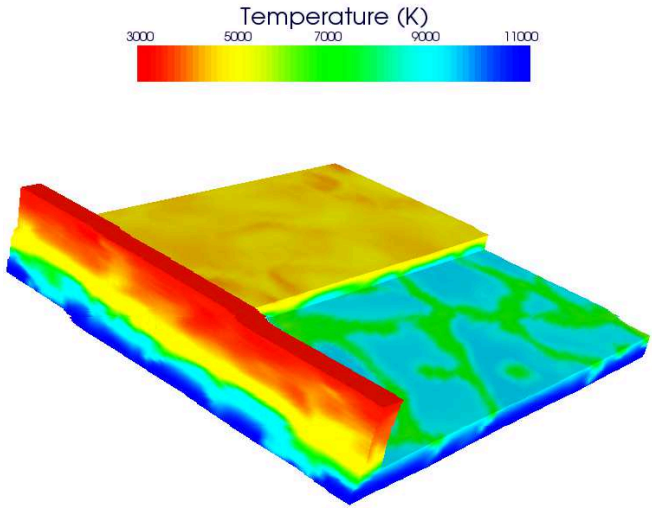}
\caption{Temperature structure of a snapshot of our 3D K-dwarf model atmosphere near the visible surface ($\mathrm{depth}=0$). The cross-section on the front wall shows the overall behavior of the granulation temperature field, which becomes cooler in the higher layers but is inhomogeneous at any given depth. Two horizontal cuts are shown; one at $\mathrm{depth}=-0.15~$Mm (predominant colors are green and blue) and another one at $\mathrm{depth}=0$ (predominant color is yellow). The granulation pattern is clearly seen at $\mathrm{depth}=-0.15~$Mm. At the visible surface, the inhomogeneities persist but to a lesser extent and the granulation pattern is much weaker.}
\label{f:k2}
\end{figure}

A snapshot of the temperature structure of our 3D model is shown in Fig.~\ref{f:k2}. Clearly, the strongest temperature contrasts occur well below the visible surface. This phenomenon was referred to as ``hidden'' granulation by \cite{nordlund90} in their study of a granulation model for $\alpha$~Cen~B (spectral type K1V) but note that the intensity fluctuations are still clearly visible on the surface of this cool star model, as illustrated by our Fig.~\ref{f:granulation_timeseries}.

\begin{figure}
\includegraphics[width=9.4cm,bb=120 380 620 845]{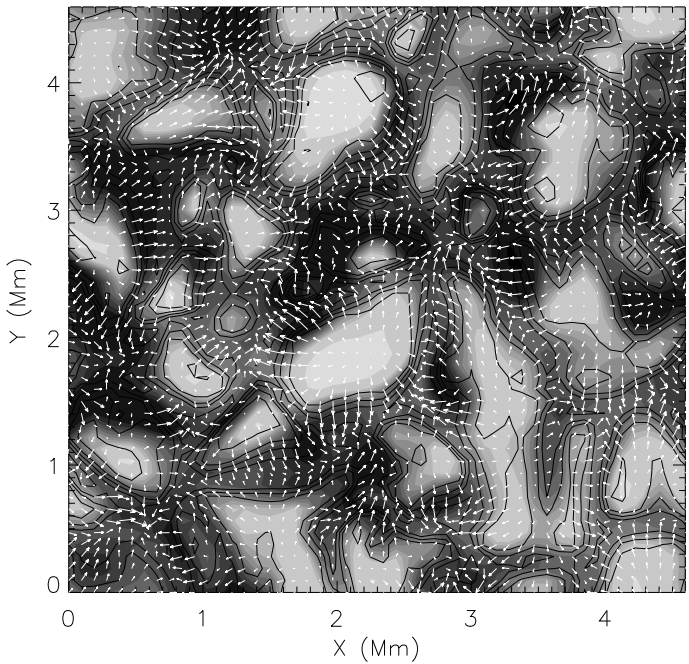}
\caption{Temperature and velocity fields of a snapshot of our 3D K-dwarf model atmosphere at the geometrical depth corresponding to 50 km below the visible surface. The velocity field is superimposed on the temperature field and is represented with solid lines (contours tracking the vertical component) and arrows (horizontal component). In this snapshot, the temperature extremes differ by about 2500~K (cf.~Fig.~\ref{f:temp3d}) while the vertical velocities range from about $-4$ to $+4$ km\,s$^{-1}$ and the horizontal component from 0 to 5.6 km\,s$^{-1}$ with a mean value of 2.1 km\,s$^{-1}$.}
\label{f:k3dtvel}
\end{figure}

\begin{figure}
\centering
\includegraphics[width=8.5cm,bb=80 375 560 700]{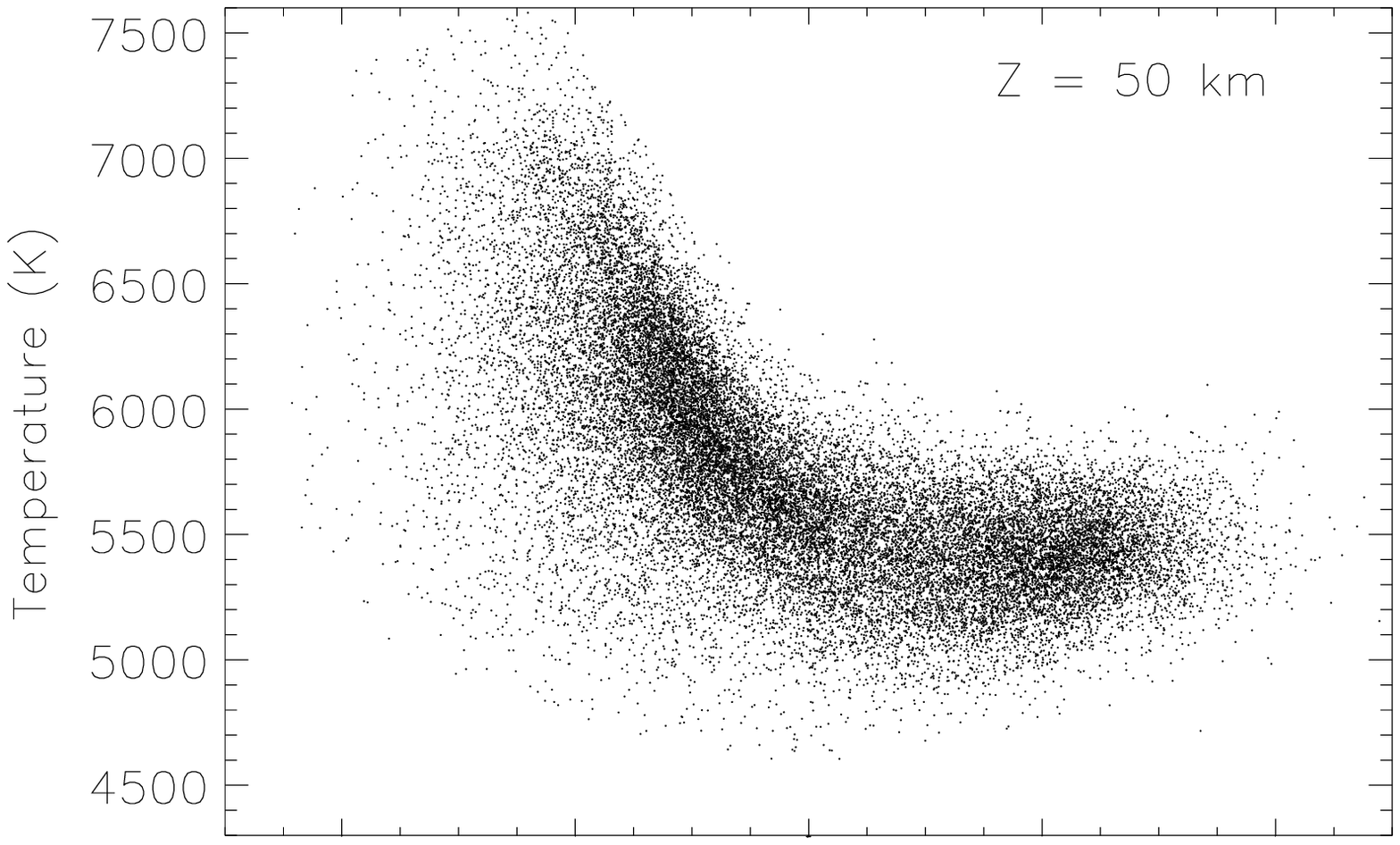}
\includegraphics[width=8.5cm,bb=80 375 560 660]{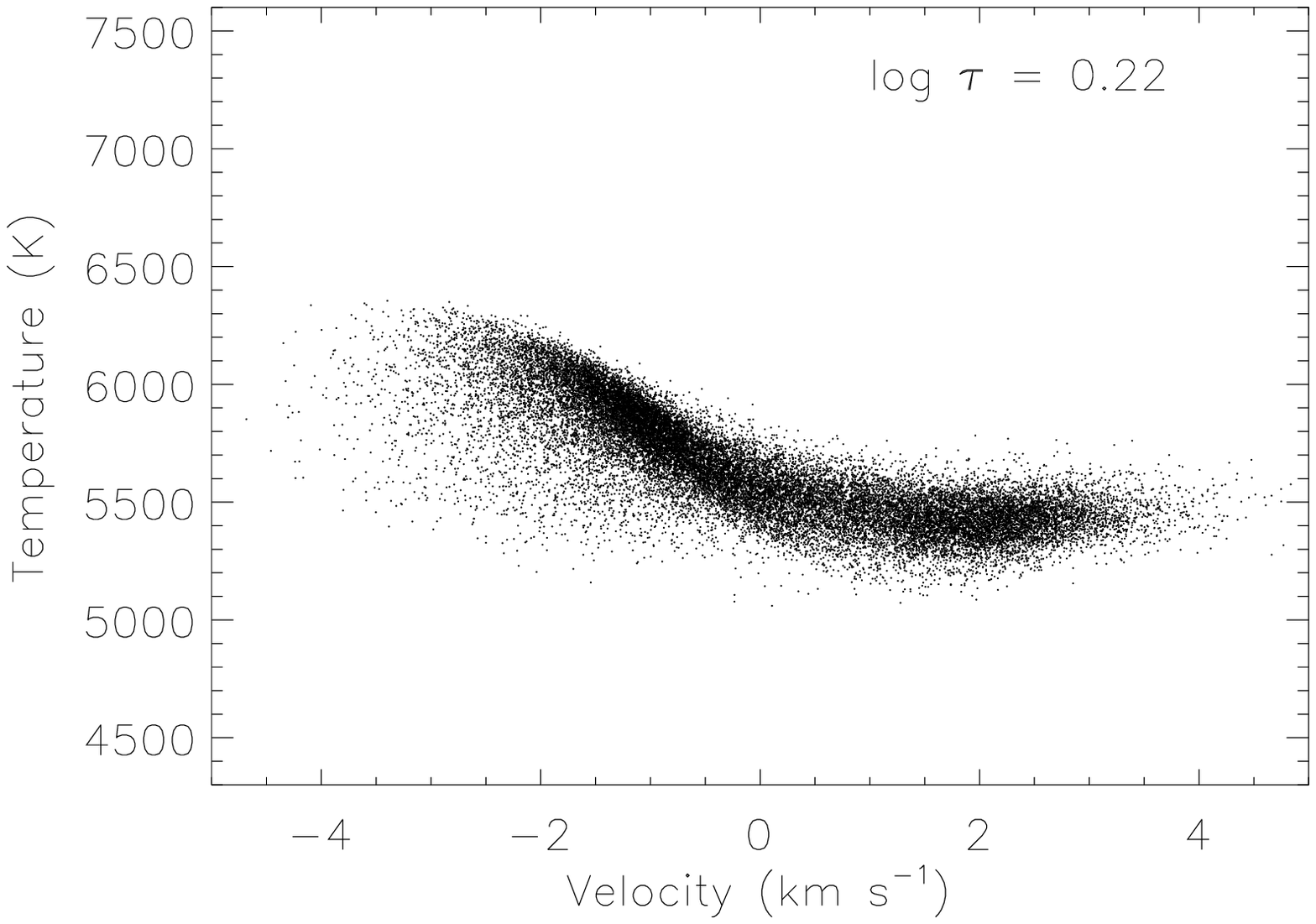}
\caption{Top panel: correlation between temperature and vertical velocity of the 3D K-dwarf model atmosphere at the geometrical depth corresponding to 50\,km below the visible surface, where the average optical depth is $<\log\tau>=0.22$. Bottom panel: as in the top panel but for local optical depth $\log\tau=0.22$.}
\label{f:tv3d}
\end{figure}

The temperature and velocity fields near the geometrical surface of zero depth are shown in Fig.~\ref{f:k3dtvel} for a given snapshot of the simulation (more precisely, this corresponds to a layer located 50\,km below the visible surface). Hot granules have the largest upward velocities and appear to be expanding from their center. The horizontal velocity field converges toward intergranular lanes, which are cooler and have the largest downward velocities. This correlation between the vertical velocity and temperature fields is illustrated also in Fig.~\ref{f:tv3d}. Note that the correlation is tighter at a given optical depth than geometrical depth. Similar results are predicted for the density and vertical velocity fields; the downflows contain higher density gas. Thus, the predicted granulation of the 3D K-dwarf model has many characteristics of the well-observed solar granulation.

\begin{figure}
\includegraphics[bb= 65 362 340 700,width=9cm]{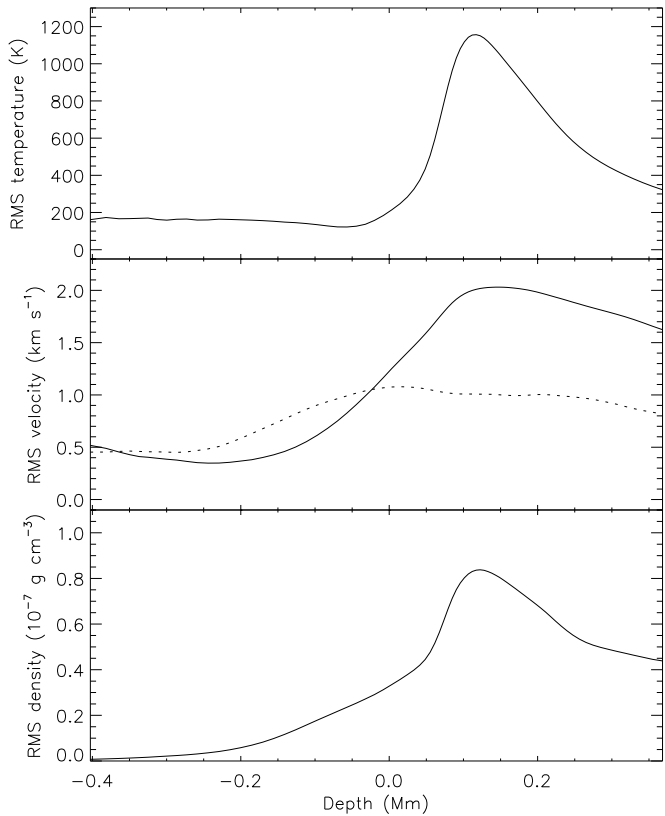}
\caption{Root mean square values of the temperature, velocity, and density fields according to our 3D model atmosphere (all snapshots included). In the middle panel, the solid line corresponds to the vertical component of the velocity while the dotted line corresponds to the horizontal component.}
\label{f:rms3d}
\end{figure}

The fluctuations of the temperature, velocity, and density fields in the 3D simulation as a function of depth are quantified by their root mean square (RMS) values. The maximum RMS values of all these parameters occur at a depth of about $+0.1$~Mm (Fig.~\ref{f:rms3d}). In this layer, the RMS values are about 1200~K, 2~km\,s$^{-1}$ (vertical component only), and $0.8\times10^{-7}$~g\,cm$^{-3}$. In the regions of line formation ($\mathrm{depth}<0$), the RMS temperature is roughly constant at about 200~K while the RMS vertical velocity decreases with height, from about 1.3~km\,s$^{-1}$ at the visible surface to 0.5~km\,s$^{-1}$ near the top of the simulation box.

Similar 3D simulations for the Sun \citep[][$\teff=5777$\,K, $\logg=4.44$, $\feh=0$]{stein98,asplund00:iron_shapes} and Procyon \citep[][$\teff=6500$\,K, $\logg=4.0$, $\feh=-0.05$]{allende02} are available for comparison. The temperature and velocity contrasts are stronger in the Procyon than in the solar model. At the visible surface ($\mathrm{depth}=0$), for example, the RMS temperature in Procyon is about 8\% compared to 4.5\% in the solar case. In the K-dwarf model, the corresponding value is even lower, about 4.0\%. The peak RMS velocities of the Procyon, Sun, and K-dwarf model are 5, 4, and 2~km\,s$^{-1}$, respectively. Thus, there is clearly an effective temperature dependence such that hotter stars experience stronger temperature and velocity contrasts. This fact has been explained by \cite{nordlund90}. Basically, the convective flux scales with the total flux so that cooler stars transport less convective energy than hotter stars. In addition, cooler stars are denser and therefore lower velocities are enough to transport the convective flux. Note also that Procyon has a lower surface gravity, which makes its photosphere less dense, thus enhancing the effect.

The granulation pattern is driven by the sharp decline in the gas temperature due to radiation losses and the subsequent lowering of the extremely temperature dependent continuous opacity. In the Procyon model, this occurs very close to the visible surface whereas it occurs a few tens to hundreds of km below it in the cooler models (about 100~km in our K-dwarf case; see Fig.~\ref{f:temp3d}). Thus, in addition to having stronger temperature and velocity contrasts, hotter 3D models such as for Procyon have their granulation pattern visible on their surfaces, a phenomenon labeled ``naked granulation'' by \cite{nordlund90}.

Although in the cooler models, such as our K-dwarf simulation, the strongest temperature contrasts occur below the visible surface, the velocity fields associated with them reach larger heights, in particular the regions of line formation. For example, Fig.~\ref{f:rms3d} clearly shows that that RMS velocity does not vanish up to 400~km above the visible surface. We therefore expect to see the granulation effects on absorption lines in K-dwarf spectra.

\section{Spectrum synthesis in 3D} \label{s:3dsynthesis}

Single spectral lines were computed with the 3D synthesis code ``{\tt lte}'' \citep[e.g.,][]{asplund00:iron_shapes}, which is described below. In Appendix~\ref{s:asset}, we compare some of these calculations with those obtained using a different 3D spectrum synthesis program, \asset\ \citep{koesterke08}, which we will use in the third part of this series \citep{kdwarfs-p3} for the synthesis of continua, molecular features, and wide spectral regions. As we show in Appendix~\ref{s:asset}, the agreement between the results obtained with the two codes is excellent.

The transfer equation was solved along several rays throughout the upper 0.8~Mm of the simulation box. The contribution from deeper layers to the emergent flux is negligible, thus justifying the use of only the upper layers in the line calculations. To improve the vertical sampling, which is necessary due to the rapid change of the local physical parameters in this region and the sensitivity of the line opacities to these quantities, the upper 0.8~Mm layers of the simulation box were interpolated to a grid of 82 depth points. In the horizontal direction, of the original $150\times150$ grid, a coarser $50\times50$ grid was adopted by eliminating two of every three grid points along the $x$ and $y$ axes, thus keeping the same geometrical extent. This has the advantage of significantly reducing computing time while keeping the results essentially unchanged \citep{asplund00:resolution}. The model used for the line calculations is therefore a $50\times50\times82$ grid.

The time interval between the snapshots of the 3D model that were saved from the original simulation is about 40 seconds. For the spectral line calculations, we used only every other saved snapshot, as tests showed that using all of them produced essentially the same results. The maximum difference in flux from this experiment was from less than 0.01\% for weak lines to about 0.04\% for the strongest lines. Tests using one every four and one every eight saved snapshots showed that this maximum flux difference increased by factors of only two (i.e., about 0.08\%) and three (0.12\%), respectively. In addition to taking into account the time dependence of the granulation phenomenon, the use of several snapshots can be interpreted as the equivalent of increased spatial coverage.

We solved the equation of radiative transfer using 8 polar ($\mu=\cos\theta$ in the standard stellar atmosphere notation) and 8 azimuthal ($\phi$) angles. The number of frequencies adopted in each case was 71, with a fine frequency spacing corresponding to 0.4~km\,s$^{-1}$. For strong lines, this frequency sampling was not enough to reach the full extent of the wings and we therefore repeated the calculation using 71 frequencies with a spacing of 1.5~km\,s$^{-1}$. We did this instead of calculating the whole wide spectral range with fine spacing to save computing time. The two results were then merged into single line profiles.

Similarly to the model atmosphere computation, LTE level populations and ionization fractions were adopted in the line calculations. Look-up tables for the equation of state and opacity for the wavelength region in which the problem spectral line is located were calculated before solving the radiative transfer. Interpolation from these tables during the spectral line synthesis instead of on-the-fly calculations reduced dramatically the computing time.

Each spectral line was computed for 3 different values of the oscillator strength ($\log gf$ and $\log gf\pm0.5$). Note that this is equivalent to computing each spectral line for three different abundances, namely $A_\mathrm{Fe}$ and $A_\mathrm{Fe}\pm0.5$~dex ($A_\mathrm{X}$ represents the number density of X relative to hydrogen in a logarithmic scale where the hydrogen abundance is defined as $A_\mathrm{H}=12$). The adopted iron abundance was solar, more precisely, $A_\mathrm{Fe}=7.45$ \citep{asplund05:solarabundances}.

We calculated line profiles for the 119 Fe~\textsc{i} and 13 Fe~\textsc{ii} lines used by \cite{ramirez07}, which are representative of the line-lists adopted in many FGK stellar abundance studies. The sources of atomic data are reliable; transition probabilities have been accurately measured in the laboratory, often by more than one group and averaged after checking for consistency between groups, while the collisional broadening parameters, in particular the van der Waals damping constants, have been obtained from recent theoretical calculations (see Sect.~4.2 in \citealt{ramirez07} for details and references). To improve the line strength coverage, we added 6 very strong \fei\ lines to this list. The atomic data adopted for these strong lines are given in Table~\ref{t:strongfei}.

\begin{table}
\caption{Strong \fei\ lines added to the line list of \cite{ramirez07} for the 3D calculation of synthetic line profiles.}
\centering
\begin{tabular}{ccccc}\hline\hline
Wavelength & $\log gf^{\mathrm{a}}$ & EP & $\sigma^{\mathrm{b}}$ & $\alpha^{\mathrm{b}}$ \\
(\AA)      &           & (eV) & (a.u.) & \\ \hline
5741.85 & $-1.67$ & 4.26 & 725 & 0.232 \\ 
6252.56 & $-1.77$ & 2.40 & 326 & 0.245 \\
6335.32 & $-2.18$ & 2.20 & 275 & 0.261 \\
6411.65 & $-0.72$ & 3.65 & 820 & 0.247 \\
6430.85 & $-1.95$ & 2.18 & 272 & 0.257 \\
6677.99 & $-1.42$ & 2.69 & 313 & 0.268 \\ \hline
\end{tabular}
\begin{list}{}{}
\item[$^{\mathrm{a}}$] Transition probabilities are from the laboratory measurements of the Oxford group \citep[e.g.,][]{blackwell76}. 
\item[$^{\mathrm{b}}$] The last two columns are the van der Waals damping constants obtained by \cite{barklem00}; $\sigma$ is the broadening cross section for an atom-perturber relative velocity $v_0=10^6$~cm~s$^{-1}$, given in atomic units ($1~\mathrm{a.u.}=2.8\times10^{-17}~\mathrm{cm}^2$), and $\alpha$ the velocity parameter, which is related to the temperature dependence of the cross-section ($\sigma\propto T^{(1-\alpha)/2}$). 
\end{list}
\label{t:strongfei}
\end{table}

The Doppler shifts introduced by the velocity field of the simulation were taken into account in the synthesis of the absorption line profiles, as well as the thermal and collisional broadening. The microturbulence and macroturbulence parameters are not needed in 3D spectrum synthesis and were therefore not used.

\section{Disk-averaged line profiles} \label{s:davprof}

\subsection{Calculation of flux profiles}

The calculation of flux from intensities is straightforward, and very accurate provided enough rays ($\mu$ and $\phi$ angles) are included in the integrations. The rotationally broadened intensity is
\[
 I(\Delta\mathrm{v},\mu,\phi,V\sin i)=\frac{1}{2\pi}\int_0^{2\pi}I(\Delta\mathrm{v}-V\sin i\sin\theta\cos\phi',\mu,\phi)\,d\phi' \nonumber
\]
\begin{equation}
\label{eq:rot1}
\end{equation}
where $V\sin i$ is the projected rotational velocity of the star, and $\phi'$ the latitude angle on the stellar disk. The frequency dependence of the intensity has been replaced here with a velocity variable $\Delta \mathrm{v}=c\Delta\nu/\nu$. The disk averaged line profile is then
\begin{equation}
F(\Delta\mathrm{v},V\sin i)=\int_0^{2\pi}\int_0^1 I(\Delta\mathrm{v},\mu,\phi,V\sin i)\,\mu\,d\mu\,d\phi\ .
\label{eq:rot2}
\end{equation}

\subsection{Bisectors and wavelength shifts of $V\sin i=0$ line profiles} \label{s:bisectorswavshiftsnobro}

\begin{figure}
\includegraphics[bb=70 380 280 975,width=6.7cm]{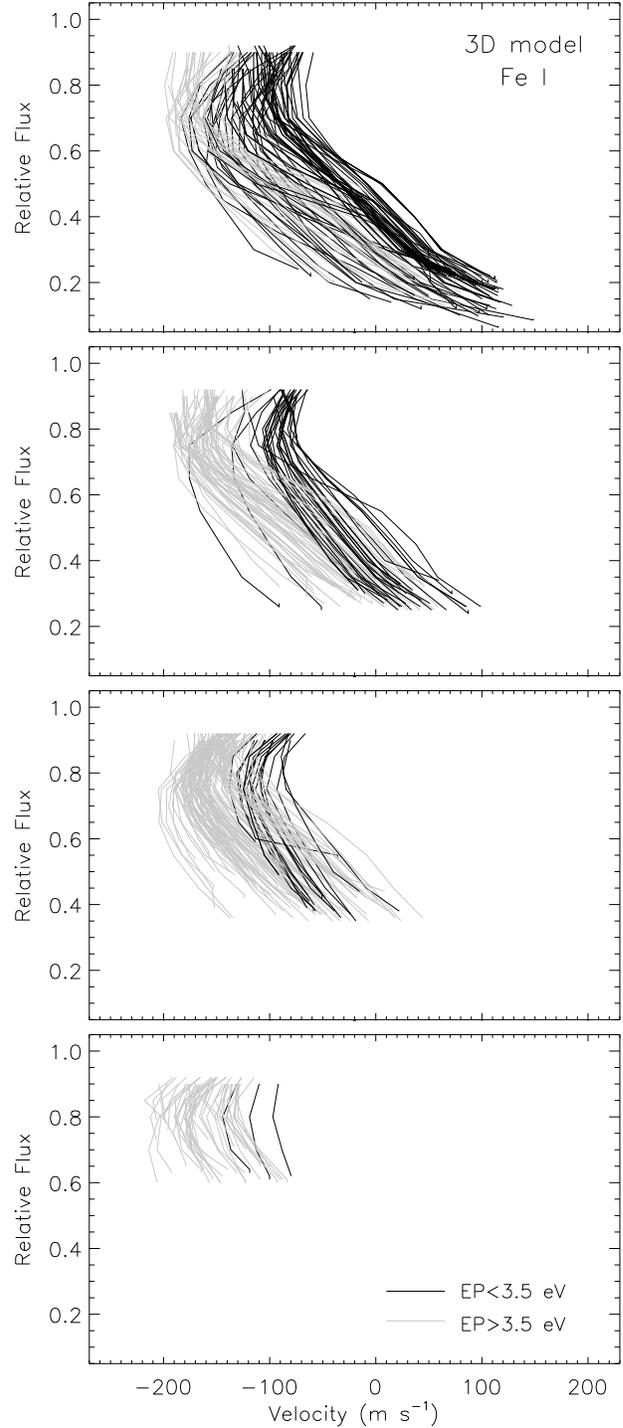}
\caption{Bisectors of the disk-averaged \fei\ line profiles calculated with our 3D model. Four groups of lines, sorted according to their line-depth, are shown in the different panels. The line bisectors have also been divided into two groups according to the value of their excitation potential (see legend on the bottom panel).}
\label{f:meanbist}
\end{figure}

We used Eqs.~\ref{eq:rot1} and \ref{eq:rot2} to determine disk-averaged line profiles adopting $V\sin i=0$. The bisectors of these line profiles (\fei\ lines only) are shown in Fig.~\ref{f:meanbist}, where they have been grouped according to line-depth and excitation potential (EP).

\begin{figure}
\includegraphics[width=9.2cm,bb=70 365 394 625]{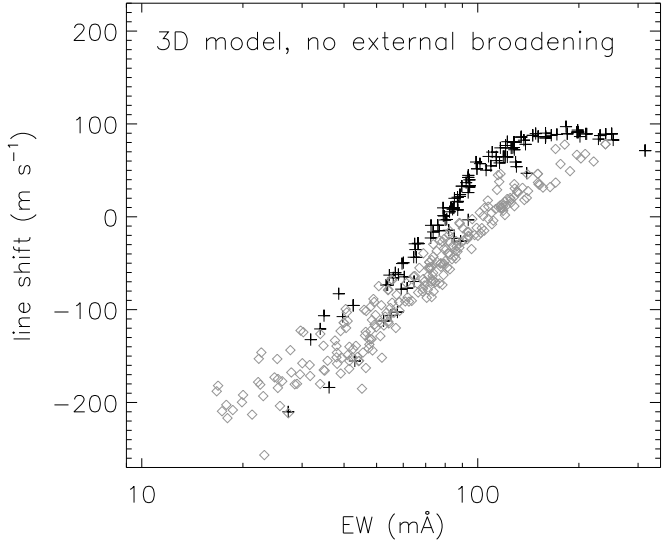}
\caption{Core wavelength shifts predicted by our K-dwarf 3D model atmosphere. The Fe~\textsc{i} lines have been grouped into low excitation potential ($\mathrm{EP}<3.5$~eV, black crosses) and high excitation potential ($\mathrm{EP}>3.5$~eV, gray diamonds) lines.}
\label{f:lshtheo}
\end{figure}

According to the 3D model, the line bisector span\footnote{We define the ``span'' as the difference in velocity between the reddest and bluest points of the line bisector.} ranges from about 250~\ms\ for the strongest \fei\ lines to about 10~\ms\ for the weakest features. The line bisectors show the characteristic C-shape attributed to granulation. The core wavelengths of the weak \fei\ features are shifted by up to about $-200$~\ms. The blueshift decreases for stronger lines and it becomes nearly zero for features of residual core flux around 0.3. Interestingly, the 3D model predicts a core wavelength \textit{redshift} for the strongest features, which amounts up to 100~\ms. This result is discussed in more detail below. The core wavelength shifts are also given as a function of the line equivalent width ($EW$) in Fig.~\ref{f:lshtheo}.

The large dispersion present between bisectors of different lines that have similar line strength, as shown in Fig.~\ref{f:meanbist}, is mostly due to the EP dependence of the core wavelength shifts. Note, for example, that lines of $EW\simeq50$\,m\AA\ have a shift that is about 30~\ms\ higher for the low EP lines compared to those that correspond to the high EP lines, as shown in Fig.~\ref{f:lshtheo}. Due to this effect, there appear to be two branches in the line shift vs.~$EW$ relations shown in Fig.~\ref{f:lshtheo}, which is, however, due to the fact that our line selection was such that few lines of intermediate EP were included; most of our Fe~\textsc{i} lines have either $\mathrm{EP}\simeq2.5$~eV or $\mathrm{EP}\simeq4.5$~eV.

When the bisectors of lines of similar strength are compared excluding the core wavelength shifts (therefore comparing only their shapes), the dispersion is reduced significantly, although a small EP dependence for the detailed shapes of the bisectors remains. Qualitatively, while the shape of the lower half of the bisector is nearly independent of EP, for a given line strength, the upper half extends more towards the blue for the high EP lines. Therefore, for a given line depth, the bisector span is slightly larger for higher EP lines.

Fig.~\ref{f:lshtheo} also shows the characteristic signature of granulation. The weakest lines, those that are formed in deep photospheric layers that experience the largest granulation effects, have the highest (in absolute value) core wavelength shifts. As the lines get stronger, and therefore the line formation depth decreases reaching higher photospheric layers with weaker granulation contrasts, the core wavelength shifts become lower. 

\begin{figure}
\includegraphics[width=9.4cm,bb=80 365 560 695]{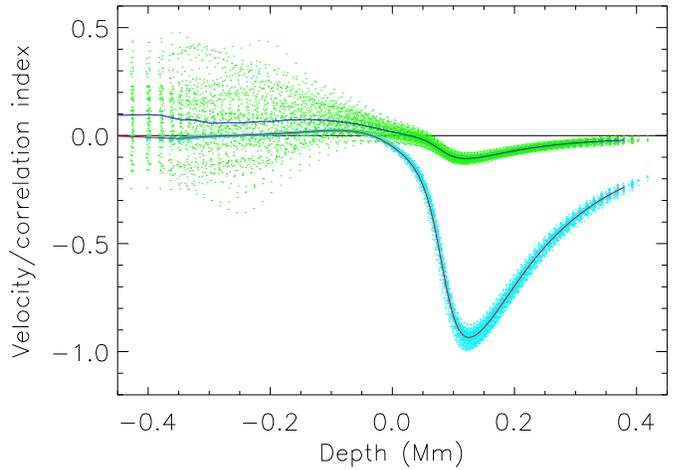}
\caption{Velocity and temperature-velocity correlation indices as a function of depth in our 3D K-dwarf model atmosphere (all 100 snapshots are shown here). The velocity index (green dots) is defined as $\overline{v_z}/\sigma(v_z)$, where $\overline{v_z}$ is the average value of the vertical velocity field at a given depth and $\sigma(v_z)$ its standard deviation. The temperature-velocity correlation index (cyan dots) is given by ${0.5\,n_g^{-1}\,\sum (v_z-\overline{v_z})(T-\overline{T})}$, where $n_g$ is the number of model grid points and $T$ the velocity field at a given depth with a mean value of $\overline{T}$. The blue and red lines correspond to these indices averaged over all snapshots.}
\label{f:vt_3dmod_kdwarf}
\end{figure}

Since the correlation between velocity and temperature fields decreases as one reaches higher atmospheric layers (see the cyan dots in Fig.~\ref{f:vt_3dmod_kdwarf}), one would expect the core wavelength shift of the strongest lines to converge towards zero. Even if the properties of some high photospheric layer are such that most of the light emitted there comes from downflows (as also suggested by Fig.~\ref{f:vt_3dmod_kdwarf}; see green dots and blue line), we would expect the redshift to decrease shortly after, provided we continue probing higher layers with stronger absorption lines. A hint of this effect is seen in Fig.~\ref{f:lshtheo}, where at $EW\simeq180$~m\AA\ the line shift, which is already a redshift, seems to start decreasing for higher $EW$ values.

In their study of line formation in solar granulation, \cite{asplund00:iron_shapes} encountered a problem for these predicted convective redshifts (see Figs.~11 and 12 in their paper). For the strongest \fei\ features, they are up to 200~\ms\ higher than the observed ones. Since the wavelengths of the solar spectrum have been accurately calibrated in an absolute sense and, in particular, no correlation between the errors in the wavelength scale and the line strength are expected in the solar atlases, this may suggest that the predicted core wavelength redshifts for the strongest lines are a numerical artifact, possibly due to deficiencies in the modeling of the outer boundary \citep{asplund00:resolution,asplund00:iron_shapes}. A test was performed to support this claim and is described in the next paragraph. Note, however, that non-LTE effects and the presence of the chromosphere in real stars may be the dominant factors in determining the absolute core wavelength shifts of those lines, which would make these particular 3D model predictions merely incomplete and not necessarily incorrect. In fact, observations of the solar photosphere show that a reversal of the granulation pattern does occur a few hundred kilometers above the visible surface \citep[e.g.,][]{janssen06}, which would explain line core wavelength redshifts. In our K-dwarf model, a weak temperature-velocity correlation (of opposite sign compared to that of the ``normal'' granulation pattern seen in continuum formation layers) is predicted at about $-0.1$\,Mm (see red line in Fig.~\ref{f:vt_3dmod_kdwarf}).

A subset of \fei\ features from our linelist were recomputed after excluding a number of upper layers (1, 2, and 4) from the simulation. Their core wavelength shifts were then calculated and compared to those obtained from the original model. We find that for \fei\ lines of $EW>100$~m\AA, the upper layers play an important role in determining the core wavelength shift. For example, for an $EW=130$\,m\AA\ line, removing 1, 2, and 4 layers increased those shifts by about 6, 13, and 24~\ms, respectively. For \fei\ lines of $EW<100$~m\AA, however, the upper layers are unimportant regarding the core wavelength shift. The differences in the core wavelengths obtained with the full model and those derived excluding upper layers have a mean value of essentially zero, with an RMS scatter of about 1 m~s$^{-1}$. These lines are therefore free from systematic errors due to the limited height of the simulation box and reliable to test the line formation of spectral features commonly used in stellar abundance determinations.

For $EW>100$\,m\AA, the core wavelength redshift diminishes as more layers are added to the line profile calculation. It is tempting to conclude from this that the redshifts of the strongest lines predicted by the 3D model will become lower if more layers are added to the simulation (although admittedly from this test alone it is clear that the correction would be insufficient to explain the +100~\ms\ predicted by the original simulation for some of the strongest \fei\ features). However, the impact of the chromosphere on these upper layers in real stars cannot be ignored, which would make our no-chromosphere model for these uppermost layers unrealistic.

Even though these predicted redshifts appear to reveal a limitation of our current 3D models, note that the line strengths are not affected in the same manner. For example, for the test described above, the corresponding change in equivalent width was between 0 and 1.5~m\AA. For the strongest features, removing 4 of the upper layers resulted in a line equivalent width lower by only about 1\%, a number that is comparable to, if not often smaller than, the typical errors in the measurement of line equivalent widths. Thus, the 3D model predictions for the \textit{strength} of \fei\ spectral features of $EW>100$~m\AA\ are still reliable, given that most of the absorption occurs in layers far from the upper boundary of our 3D model.

\begin{figure}
\includegraphics[bb=80 810 280 975,width=6.8cm]{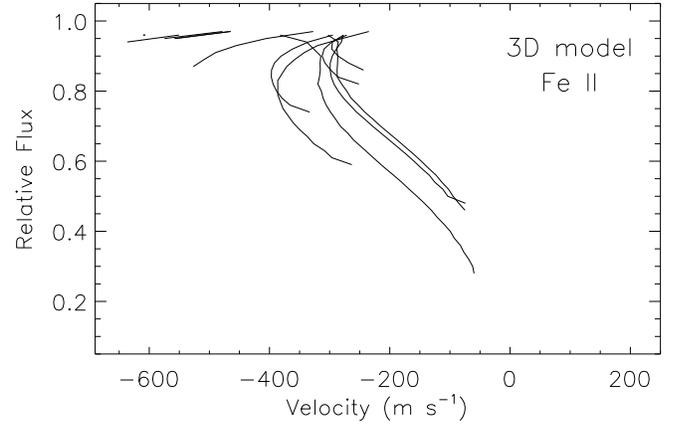}
\caption{Bisectors of the disk-averaged \feii\ line profiles calculated with our 3D model and assuming $V\sin i=0$.}
\label{f:meanbist2}
\end{figure}

\feii\ lines suffer from stronger granulation effects. Their bisectors span from about 100 to slightly more than 200~\ms\ while the core wavelength blueshifts of the weak lines reach up to about $-600$~\ms\ (Fig.~\ref{f:meanbist2}). For $EW$ values between 15 and 40~m\AA, which is the only $EW$ range where our \fei\ and \feii\ lines overlap, the convective blueshifts of \feii\ lines are, on average, about 100~m~s$^{-1}$ bluer (i.e., more negative) than those of \fei\ lines. This is due, most likely, to the deeper formation depths of \feii\ lines. The \feii\ number density increases with depth as higher temperatures are required to ionize the neutral iron atom. At these deeper layers, the intensity and velocity fields are strongly correlated. Although it would be ideal to look for the granulation signatures using \feii\ lines, in real K-dwarf spectra only a small number of them are available for accurate line profile measurements.

\subsection{Rotationally broadened profiles} \label{s:rotation}

\begin{figure}
\includegraphics[bb=90 370 470 625,width=9cm]{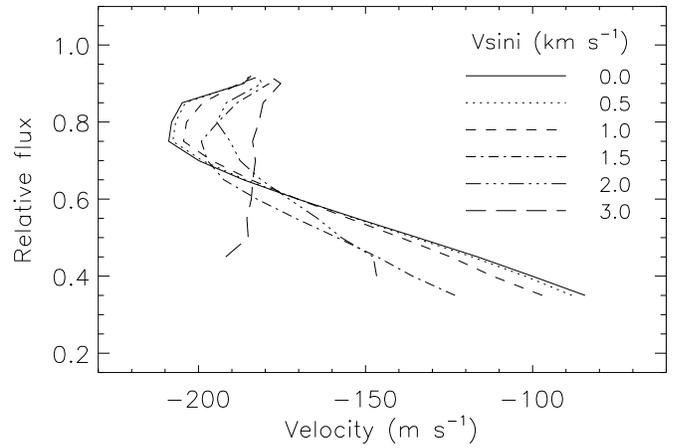}
\caption{Bisectors of the theoretical 5228.4 Fe~\textsc{i} line computed for several values of the projected rotational velocity $V\sin i$ (see legend).}
\label{f:testrot3d}
\end{figure}

\begin{figure}
\includegraphics[bb=90 370 470 625,width=9cm]{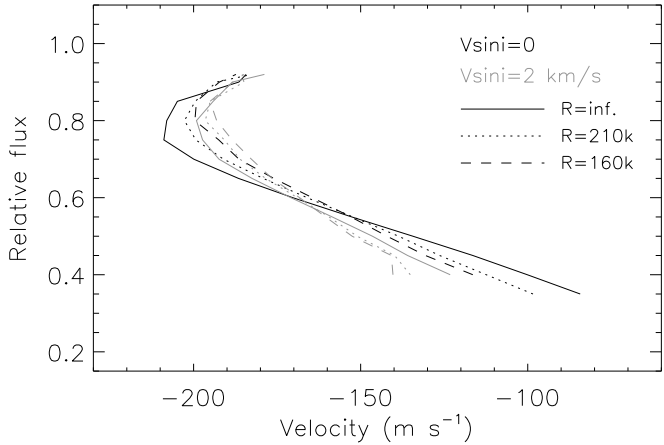}
\caption{Bisectors of the theoretical 5228.4 Fe~\textsc{i} line computed for $V\sin i=0$ (black lines) and 2~km~s$^{-1}$ (gray lines), and three values of the resolving power: infinite (solid lines), $R=210,000$ (dotted lines), and $R=160,000$ (dashed lines). The instrumental profile was assumed to be a Gaussian for this test.}
\label{f:testres}
\end{figure}

To explore the effect of the projected rotational velocity on line profiles affected by granulation, we computed a few spectral lines using Eqs.~\ref{eq:rot1} and \ref{eq:rot2} with several non-zero $V\sin i$ values and compared the resultant line bisectors. One typical example of this exercise is shown in Fig.~\ref{f:testrot3d}.

Between $V\sin i=0$ and 1~km\,s$^{-1}$, the shape of the line bisector is only slightly affected by stellar rotation and it therefore retains most of the granulation signatures. On the other hand, for projected rotational velocities greater than 1~km\,s$^{-1}$, the effects of stellar rotation must be taken into account to properly characterize them. In general, the effect of a non-zero $V\sin i$ is to make the line more symmetric. As one might naively expect, additional symmetric broadening dilutes the line asymmetries.

The effect of the rotational velocity on the convective blueshifts is also illustrated in Fig.~\ref{f:testrot3d}. Interestingly, although the line profiles become more symmetric for higher $V\sin i$ values, the absolute value of the convective blueshift increases. This is a confirmation of the ``rotation effect'' discussed by \cite{gray85} and \cite{gray86} on the basis of their numerical experiments using two component granulation models. The distribution of Doppler shifts for the granules in a $V\sin i=0$ star extends from a maximum blueshift value corresponding to granules in the disk center to zero for the limb. Since the regions observed near the limb correspond to larger areas due to projection effects, the distribution shows an increase towards lower blueshifts, peaks at a value close to the line-center convective blueshift, and vanishes at zero velocity. A large $V\sin i$ star redistributes the observed Doppler shifts from the granules in areas near the limb symmetrically about zero. Since the distribution does not change for areas near the disk center, the entire disk distribution will now have a red tail but a peak extending further towards the blue, thus producing the rotation effect. Note that this increase in the convective blueshift is an external effect that is not associated with the strength of the granulation inhomogeneities. Rotation is not taken into account in the computation of the model atmosphere, only in the calculation of disk-averaged line profiles from the emergent intensities.

\subsection{Impact of instrumental imperfections}

The observed spectral line profiles have an additional external broadening due to the finite value of the spectral resolution of the spectrograph. In our case (see \citetalias{kdwarfs-p1} for details) we found that the spectral resolution ($R=\lambda/\Delta\lambda$) was not constant but varied between 160,000 and 210,000 among our spectra. In addition, we found that the shape of the instrumental profile was slightly asymmetric. Here we explore the impact that these instrumental imperfections have on the theoretical line profiles.

\subsubsection{Variable resolution}

In order to explore the effects of a variable spectral resolution on the theoretical line profiles, we performed tests convolving the disk-averaged, rotationally broadened line profiles with Gaussian instrumental profiles of $\mathrm{FWHM}=\lambda/R$. Although the actual instrumental profile of our observations was not Gaussian, its overall shape remained nearly constant (its FWHM, however, did change with time). We performed this test with Gaussians to avoid the complications introduced by non-symmetric profiles. The latter is discussed below. We performed the test for two values of the projected rotational velocity: $\vsini=0$ and 2.0~km\,s$^{-1}$. The result is shown in Fig.~\ref{f:testres}

As it is clear from inspection of Fig.~\ref{f:testres}, the effects of a variable resolving power (between 160,000 and 210,000) on measurements of line asymmetries are relatively small, specially if the profiles have already been rotationally broadened using typical $V\sin i$ values for K-dwarfs. Lower resolution makes the lines more symmetric but, in our range of resolving power and typical values of projected rotational velocity, the asymmetry is reduced, on average, by only a few meters per second.

Note that lowering the spectral resolution using Gaussian instrumental profiles results in an increased core wavelength blueshift. This is due to the fact that the lines are asymmetric and the asymmetry is such that the blue wing extends more than what is expected for a symmetric profile. In the ``blurring'' of the spectral line, this extended wing contributes more to the shallowing of the line core than the red wing, resulting in an additional core wavelength blueshift.

\begin{figure}
\centering
\includegraphics[width=8.8cm,bb=65 400 475 550]{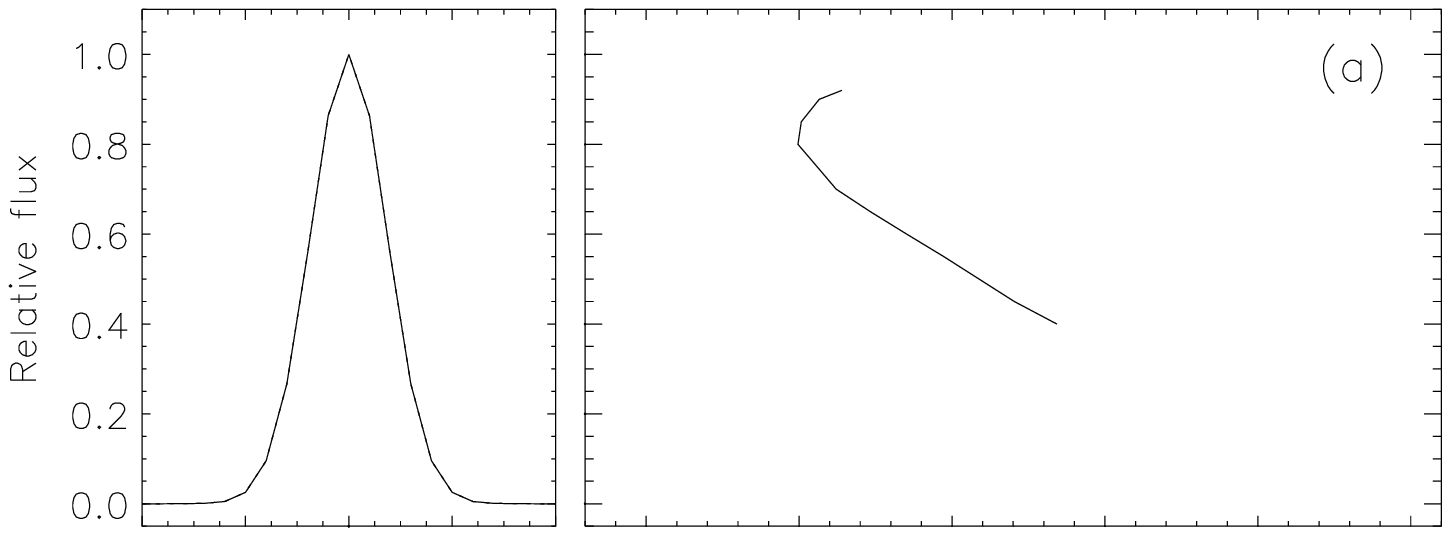}
\includegraphics[width=8.8cm,bb=65 400 475 550]{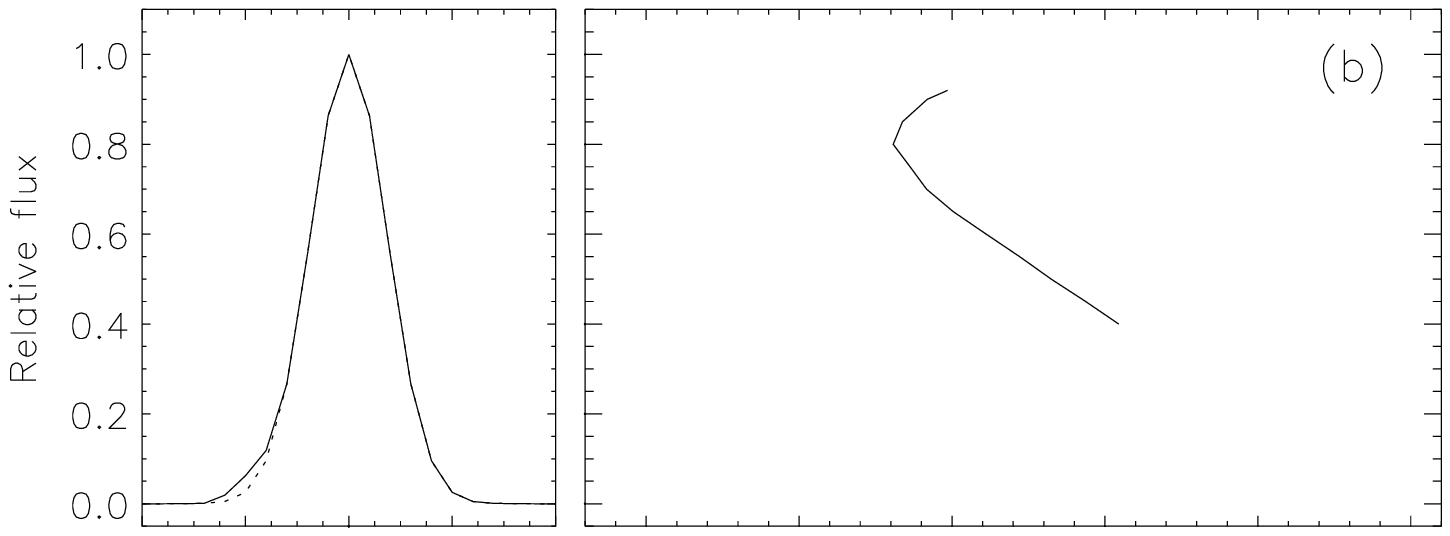}
\includegraphics[width=8.8cm,bb=65 400 475 550]{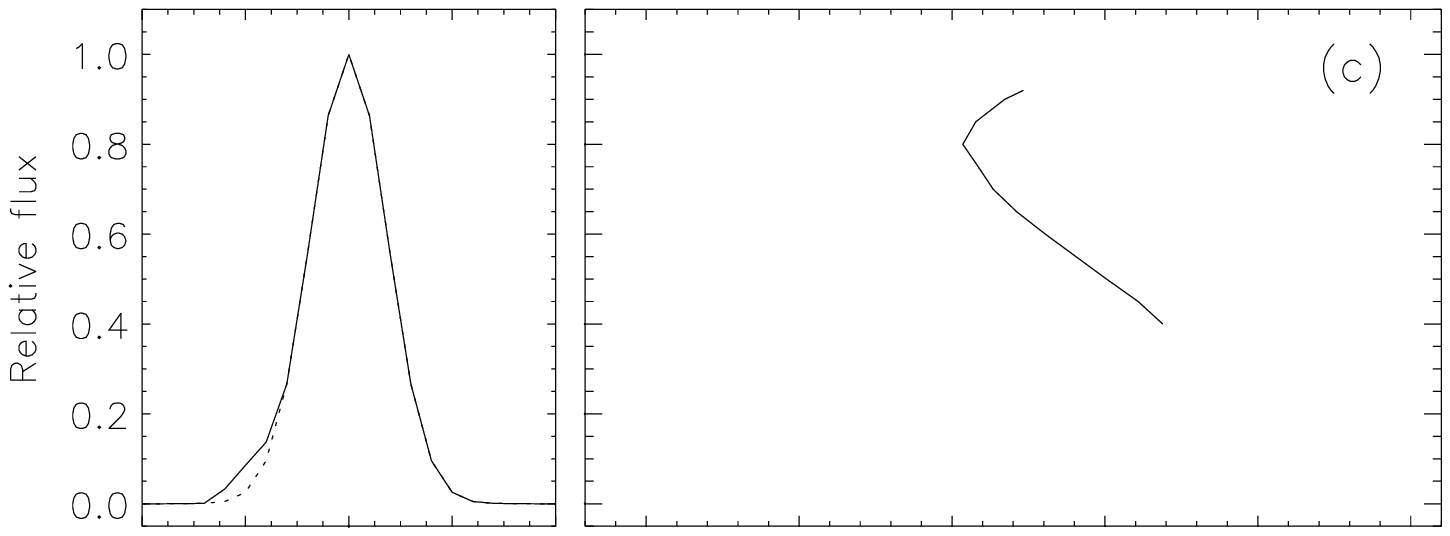}
\includegraphics[width=8.8cm,bb=65 400 475 550]{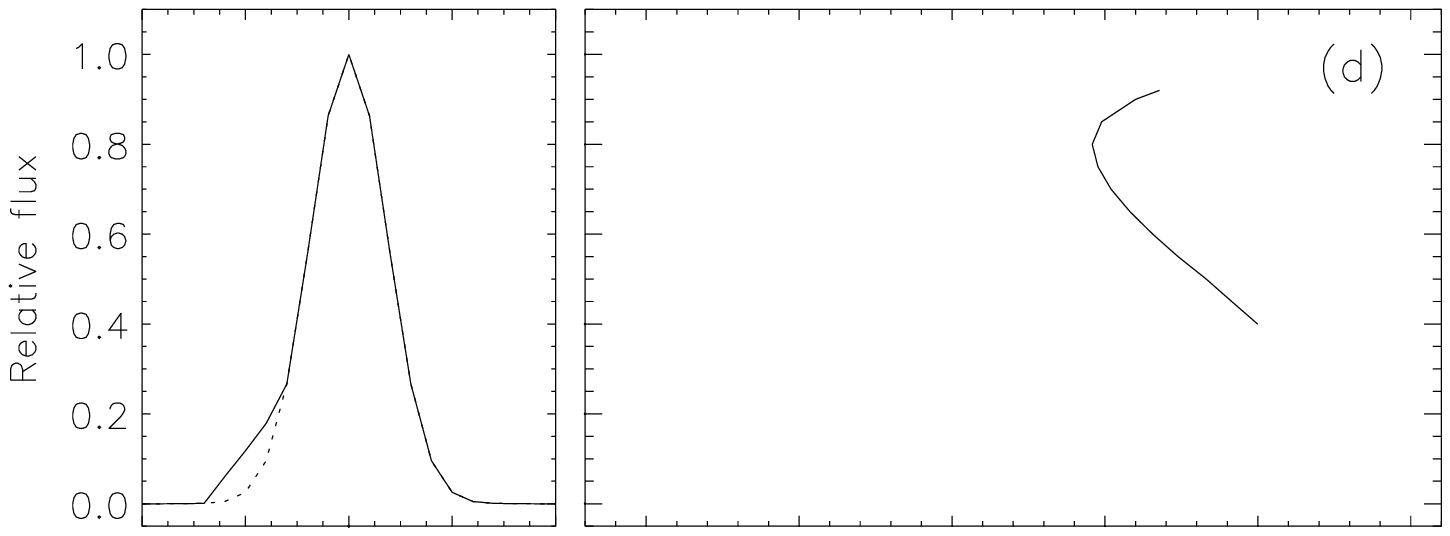}
\includegraphics[width=8.8cm,bb=65 400 475 550]{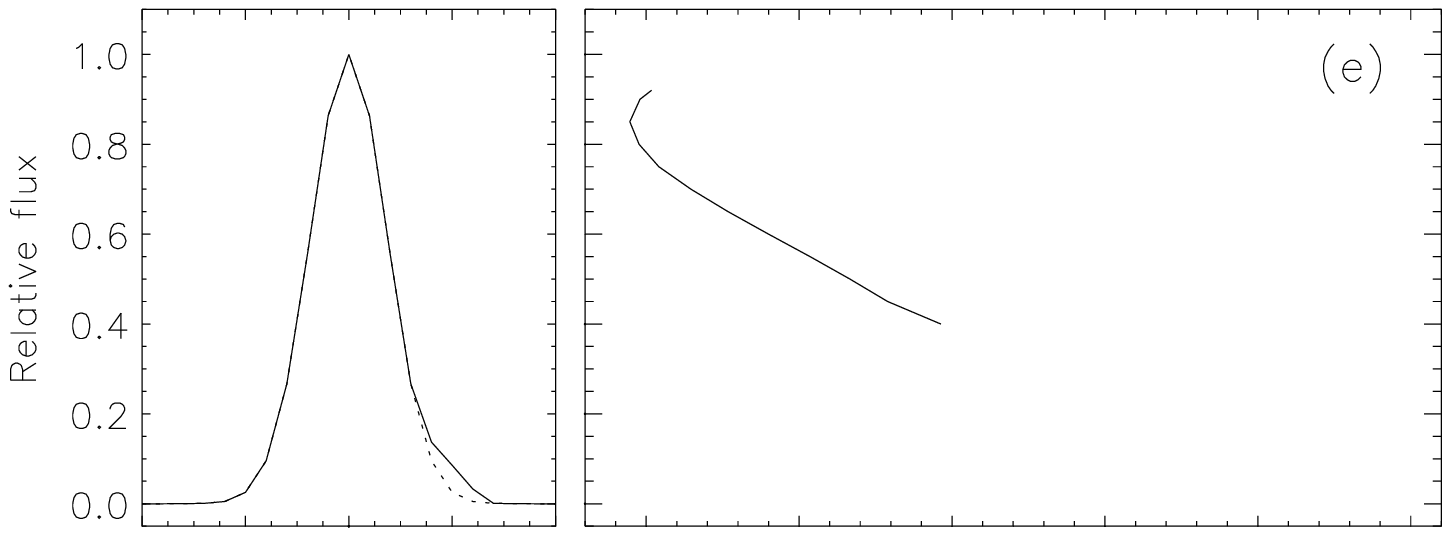}
\includegraphics[width=8.8cm,bb=65 360 475 550]{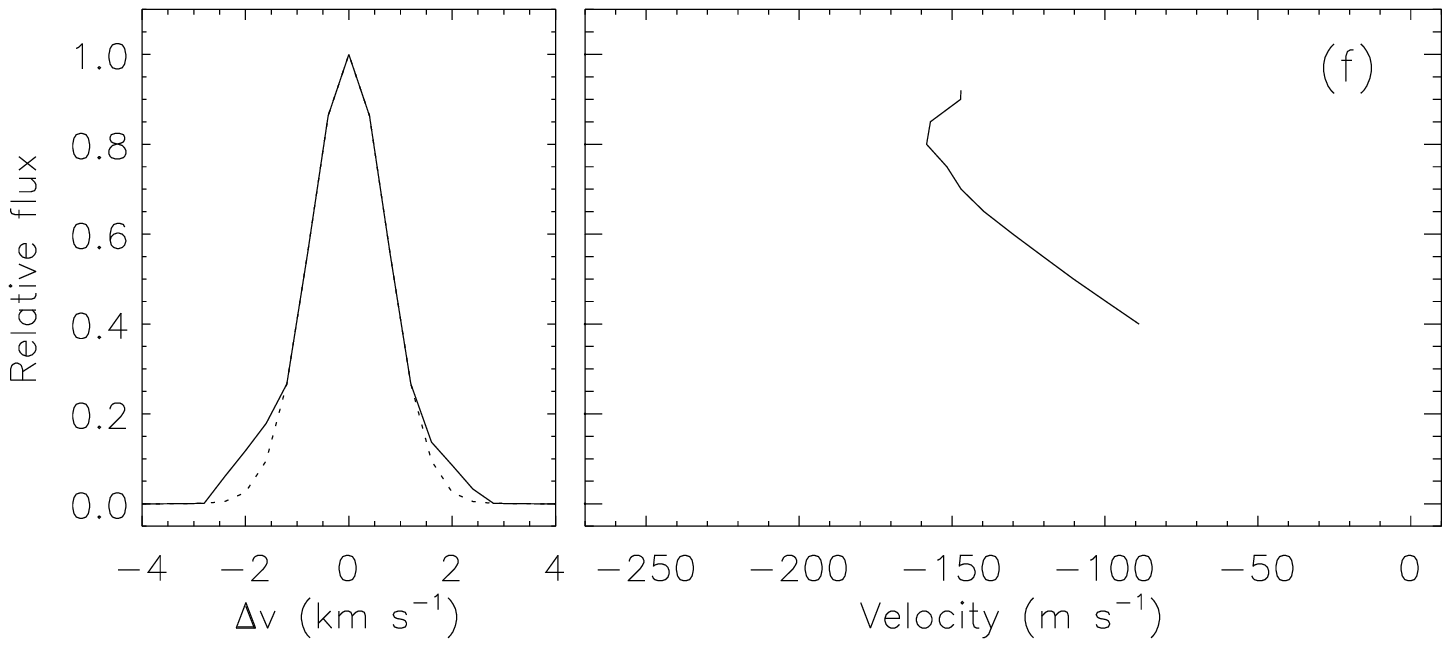}
\caption{Effect of an asymmetric instrumental profile (left panels) on the shape and core wavelength shift of the line bisector (right panels). The 5228.4\,\AA\ \fei\ line was used here to illustrate the effect. The top panel (a) shows the case of a Gaussian instrumental profile.}
\label{f:testres_show}
\end{figure}

\subsubsection{Asymmetric instrumental profile} \label{s:nogauss}

In \citetalias{kdwarfs-p1}, we showed that the instrumental profile of our observations was very similar to a Gaussian profile but not a perfect match. The largest departure from Gaussianity was observed on the blue wing, where a small bump was present and affects all our data.

Fig.~\ref{f:testres_show} shows how an asymmetric instrumental profile affects the shape of a line bisector as well as its core wavelength shift. This test was made generating a Gaussian profile and introducing small perturbations instead of using the real instrumental profile for practical reasons; the shape of the actual instrumental profile is not as smooth as a Gaussian function due to the noise of the ThAr exposures used to determine it (see Sect.~3.5 in \citetalias{kdwarfs-p1} for details) and it is not as trivial as the Gaussian case to deconvolve a non-smooth profile to take into account thermal broadening, which represents about 5\,\% of the total broadening of the Th lines (we do use the real instrumental profile later for the comparison with our observations).

Compared to the Gaussian case, our slightly asymmetric instrumental profile affects both the line bisector span and core wavelength shift of the line profiles formed in stellar granulation. A blue bump on the profile reduces the span and shifts the entire line towards the red by an amount that depends on the size of the bump, as seen by the sequence (b),(c),(d) in Fig.~\ref{f:testres_show}. The size of each of these bumps is such that the area under the instrumental profile compared to the Gaussian case is 1.6 (a), 2.8 (b), and 5\,\% (c) higher while the shifts are about 20, 40, and 70\,\ms, respectively. Similarly, a red bump that produces a 2.8\,\% departure from Gaussianity increases the span and introduces an extra blueshift of about 30~\ms, as shown by case (e). Note, however, that the instrumental profile of our observations does not show a red bump alone. Only for one of our observing runs, both a blue and a red bump were present. Panel (f) in Fig.~\ref{f:testres_show} approximately corresponds to that case (here the departure from Gaussianity corresponds to 7.8\,\% of covered area difference). The instrumental profile of our observations varied with time, but the relative departure from Gaussianity was nearly time-independent. Cases (b), (c), and (f) in Fig.~\ref{f:testres_show} fairly represent our data, with (c) corresponding to the bulk of them (cf.~Fig.~4 in \citetalias{kdwarfs-p1}).

Thus, we find that after convolving the predicted line profiles with our asymmetric instrumental profile (represented by cases b, c, and f in Fig.~\ref{f:testres_show}), the span of the line bisector is somewhat affected (it is reduced by about 10~\ms) while the core wavelength shift changes significantly (about 20--40~\ms\ towards the red) compared to the Gaussian approximation.

\begin{figure}
\includegraphics[bb=54 380 337 860,width=8.5cm]{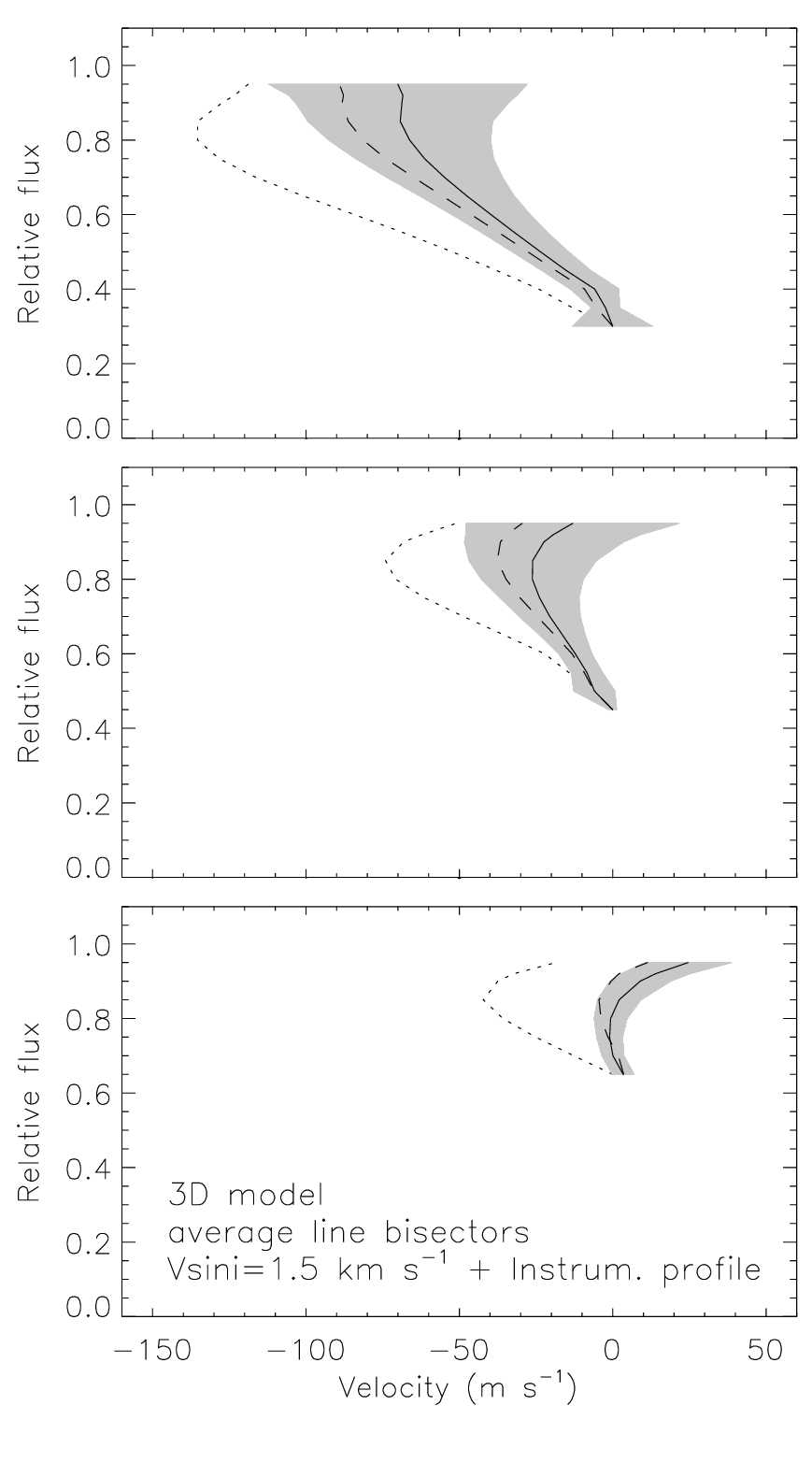}
\caption{Solid lines: mean line bisectors obtained from the theoretical line profiles computed with our 3D model and adopting $V\sin i=1.5$~km\,s$^{-1}$ and an (asymmetric) instrumental profile corresponding to our observations of the reference star HIP~86400. The shaded areas correspond to the 1-$\sigma$ dispersion of the theoretical line-by-line bisectors. Each bisector has been forced to have zero velocity at line center. The dotted lines show the mean theoretical line bisectors assuming $V\sin i=0$ and no instrumental broadening while the dashed line shows the same mean bisectors but adopting a Gaussian instrumental profile of $R\simeq180,000$.}
\label{f:mbisth3dx}
\end{figure}

\begin{figure}
\includegraphics[width=9.2cm,bb=70 365 394 623]{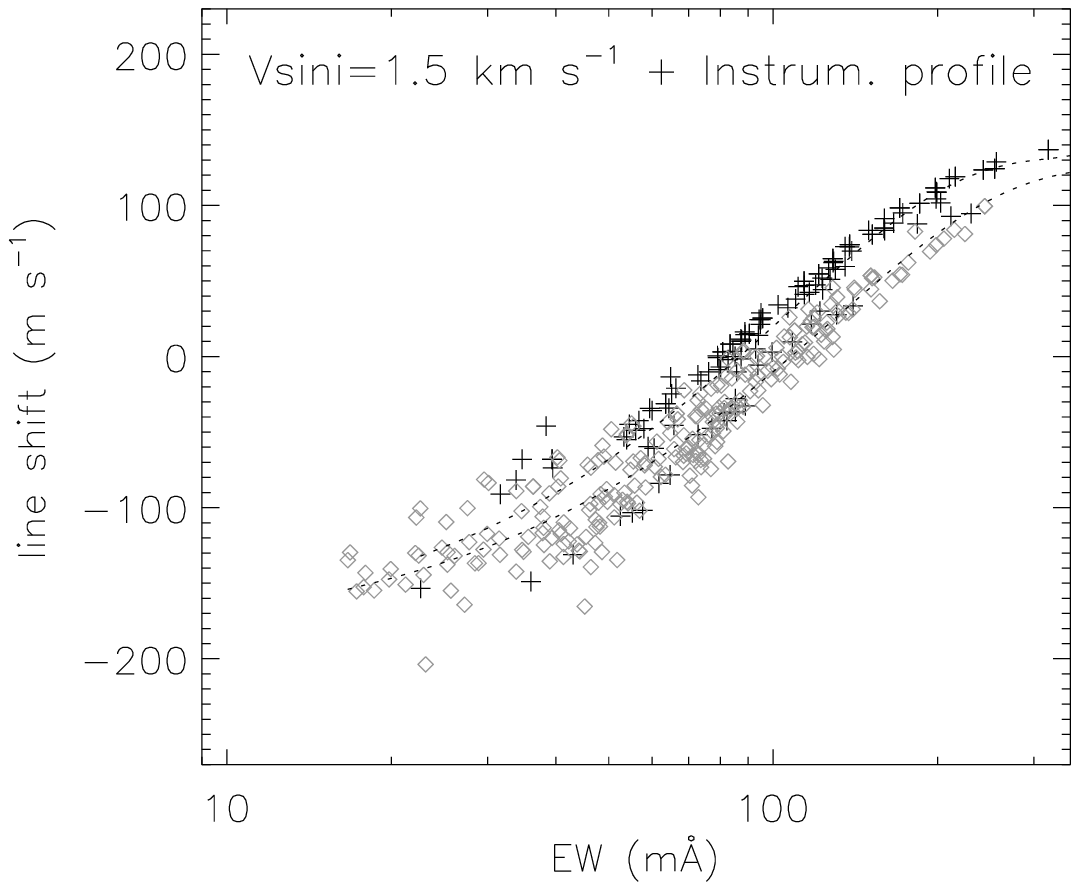}
\caption{Core wavelength shifts obtained from the theoretical line profiles computed with our 3D model and adopting $V\sin i=1.5$~km\,s$^{-1}$ and an (asymmetric) instrumental profile corresponding to our observations of the reference star HIP~86400. The Fe~\textsc{i} lines have been grouped into low excitation potential ($\mathrm{EP}<3.5$~eV, black crosses) and high excitation potential ($\mathrm{EP}>3.5$~eV, gray diamonds) lines. The dotted lines are cubic fits to the predicted line shifts of the two EP groups.}
\label{f:lshtheox}
\end{figure}

\subsection{Reference mean theoretical line bisectors and wavelength shift vs. $EW$ relation} \label{s:theoretical}

For the comparison with observed data, the line profiles predicted by our 3D model should be convolved with appropriate values of rotational and instrumental profiles. Given that these are relatively fine effects for the sample at hand and the fact that only one star in the sample (HIP~86400) has fundamental parameters very close to those of the 3D model, we define a reference case of $V\sin i=1.5$~km\,s$^{-1}$ (roughly the projected rotational velocity of HIP~86400, as shown in Appendix~\ref{s:vsini86}) and an instrumental profile that corresponds to the observing run when most of the data for HIP~86400 were acquired (see \citetalias{kdwarfs-p1}). Hereafter, we will refer to the theoretical line bisectors and wavelength shifts as those that correspond to this particular choice of external broadening parameters.

The line bisector is an excellent probe of line formation in stellar granulation because at each flux depth its value is sensitive to the inhomogeneities of a certain range of photospheric layers. We thus expect the shapes of bisectors of lines of a given strength to be similar. There will be, of course, differences due to dissimilar transition properties such as oscillator strength, excitation potential (cf. Fig.~\ref{f:meanbist}), collisional broadening, etc. Nonetheless, grouping lines of similar strength to obtain mean theoretical line bisectors is necessary in the context of this work. In \citetalias{kdwarfs-p1} we showed that measuring individual line bisectors is a very difficult task and the results very noisy, but averaging bisectors of many lines of similar line strength resulted in very robust mean line bisectors. We thus follow the same procedure with the theoretical results.

The mean theoretical line bisectors for Fe~\textsc{i} lines with residual core flux around 0.30, 0.45, and 0.65 are shown in Fig.~\ref{f:mbisth3dx}. The error bars shown there (shaded areas) correspond to the 1-$\sigma$ standard deviation from the mean, which is the \textit{intrinsic} dispersion that we would expect from comparison with observed lines. For clarity, Fig.~\ref{f:mbisth3dx} shows the line bisectors measured with respect to the line center. The core wavelength shifts are shown separately in Fig.~\ref{f:lshtheox}. We perform this separation because the observed data is treated in that way to avoid introducing the error of the rest core wavelengths in the determination of the shapes of the mean line bisectors.

For comparison, Fig.~\ref{f:mbisth3dx} also shows the mean line bisectors that are obtained from the $V\sin i=0$ profiles without instrumental broadening (dotted lines) and also those obtained from the  $V\sin i=1.5$~km\,s$^{-1}$ profiles convolved with a Gaussian instrumental profiles of FWHM equal to that of the real instrumental profile ($R\simeq180,000$; dashed lines). Clearly, the effect of the projected rotational velocity on the mean bisector shape is more important than that of the asymmetry of the instrumental profile, although the impact of the latter is not negligible, in particular for the regions near the continuum.

The core wavelength shift vs.~$EW$ relation that results from the line profiles calculated with our 3D model, after convolution with the reference rotational and instrumental profiles, is shown in Fig.~\ref{f:lshtheox}. This figure should be compared to Fig.~\ref{f:lshtheo}, which shows the corresponding result for $V\sin i=0$ and no instrumental broadening. In addition to vary slower, the reference line shift vs.~$EW$ relation (Fig.~\ref{f:lshtheox}) is shifted upwards due to the effects of the asymmetric instrumental profile (Sect.~\ref{s:nogauss}), which in our case is more important than the rotation effect that shifts the lines towards the blue instead (Sect.~\ref{s:rotation}).

In Fig.~\ref{f:lshtheox}, the lines have been separated into high ($\mathrm{EP}>3.5$~eV) and low ($\mathrm{EP}<3.5$~eV) excitation potential lines, as was done in Sect.~\ref{s:bisectorswavshiftsnobro}. Two independent cubic polynomial fits were made, one for each group, and are shown with the dotted lines in Fig.~\ref{f:lshtheox}. These fits are used below for the comparison with the observed data.

\section{Comparison to observations} \label{s:comparison}

\begin{figure}
\includegraphics[bb=90 380 280 765,width=9cm]{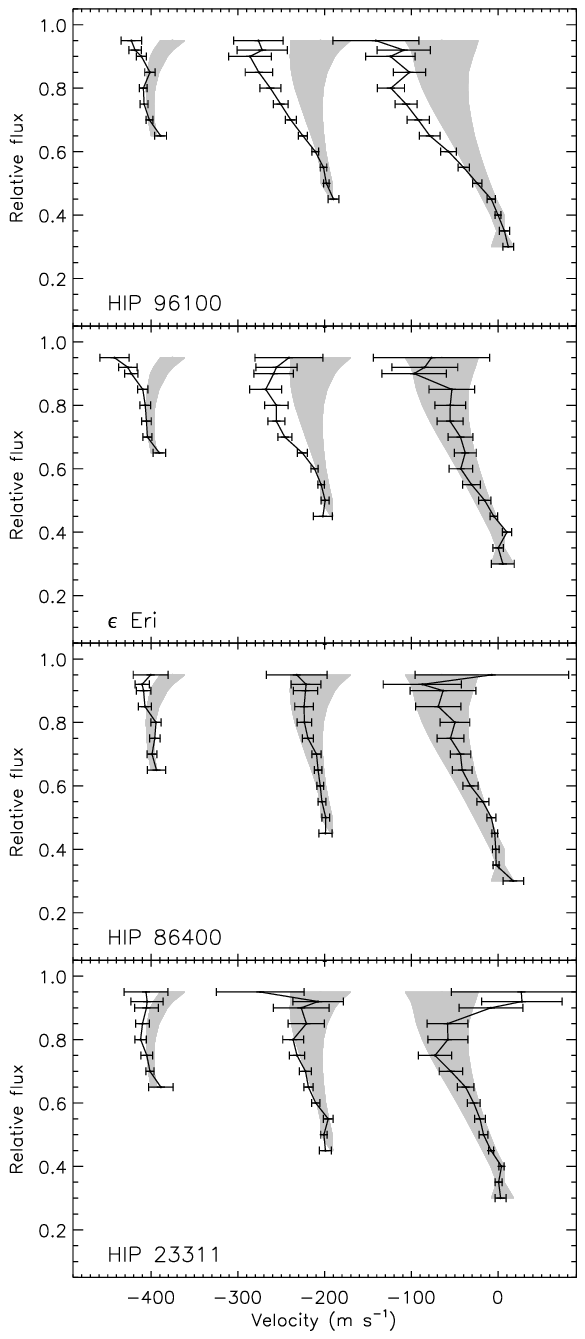}
\caption{Comparison of theoretical and observed mean \fei\ line bisectors for four K-dwarfs. These bisectors are measured with respect to the line core wavelengths, arbitrary shifts of $-200$ and $-400$~m~s$^{-1}$ have been applied to the weaker lines for clarity. The theoretical bisectors are represented by the gray shaded areas; their extent corresponds to the 1-$\sigma$ intrinsic scatter that results from the averaging of several \fei\ line profiles. The solid lines with error bars correspond to the observations (see \citetalias{kdwarfs-p1} for details).}
\label{f:bis_theo_obs}
\end{figure}

\begin{figure}
\includegraphics[bb=85 380 280 765,width=9.3cm]{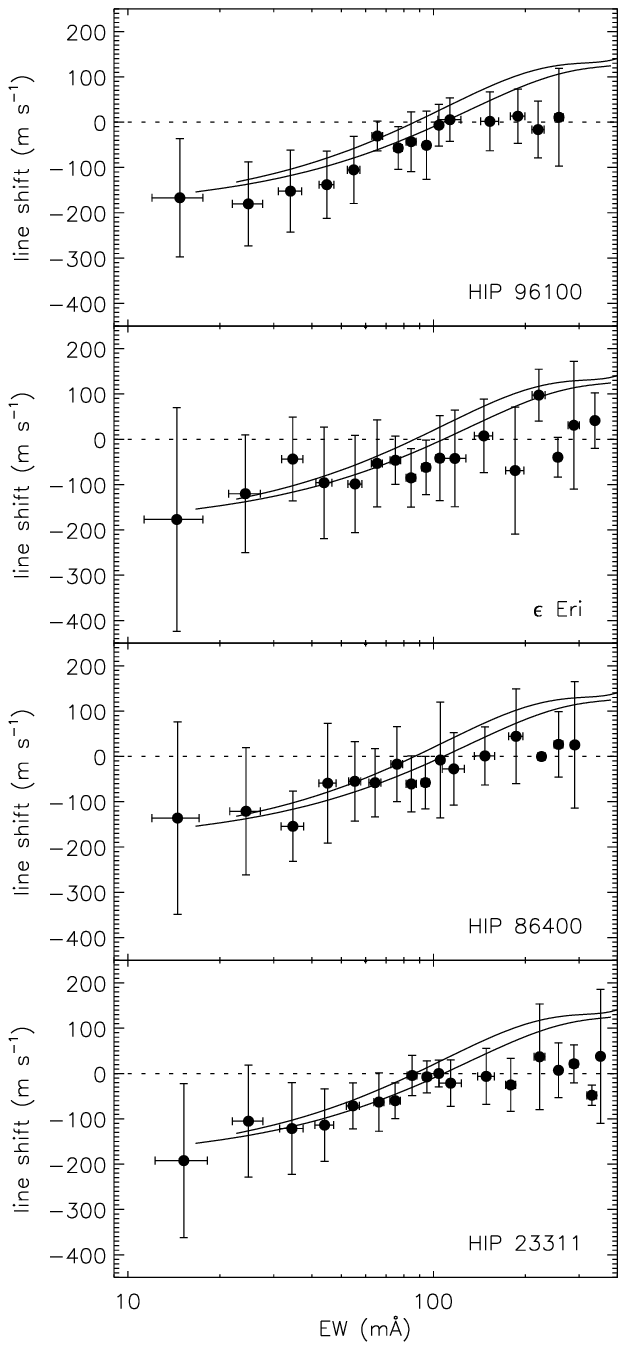}
\caption{Comparison of theoretical and observed line core wavelength shift vs.~$EW$ relation for four K-dwarfs. The observed data (filled circles with error bars) represent average values in bins of equivalent width and have been shifted so that the average velocity of the strongest lines ($EW\gtrsim100$\,m\AA) is zero (see \citetalias{kdwarfs-p1} for details). The solid lines correspond to the predictions of the 3D model (cf.~Fig~\ref{f:lshtheox}). The two solid lines correspond to the relations obtained for low and high EP lines, as explained in Sect.~\ref{s:theoretical}. Note that the $x$-axis scale is logarithmic. The dotted line corresponds to zero line core wavelength shift.}
\label{f:lshmk_theo}
\end{figure}

In \citetalias{kdwarfs-p1}, the very high spectral resolution, high signal-to-noise ratio spectra that we obtained for nine bright K-type dwarf stars were described in detail. Here those data are used to compare the predictions of our hydrodynamic model atmosphere and 3D spectrum synthesis with the observed shapes of \fei\ absorption line profiles and thus verify the accuracy of the model. In Figs.~\ref{f:bis_theo_obs} and \ref{f:lshmk_theo} we show the comparison of bisectors and wavelength shifts, respectively, for four of our sample stars.

\subsection{Line bisectors} \label{s:linebis}

In general, the basic observed shapes of the line bisectors are well reproduced by the model. Quantitatively, however, there are measurable differences. In Fig.~\ref{f:bis_theo_obs}, three average spectral line bisectors are shown in each panel. The one representing strong (weak) lines was obtained with features of line depth between 0.25 and 0.40 (0.60 and 0.80), which roughly corresponds to $EW=80-140$\,m\AA\ ($20-40$\,m\AA). The intermediate strength bisector corresponds to all other lines in between (i.e., line depth between 0.4 and 0.6 and $EW=40-80$\,m\AA).

The mean line bisectors of HIP~96100 ($\teff\simeq5220$\,K) have clearly larger span than those predicted by the 3D model (see the top panel in Fig.~\ref{f:bis_theo_obs}). There are three reasons that can explain this. First, the effective temperature of this star is about 400~K hotter than the 3D model, second, its metallicity is lower than solar ($\feh=-0.22$), and, finally, its projected rotational velocity is the lowest among our sample stars ($V\sin i\simeq0.8$~km~s$^{-1}$).\footnote{For more details and references on the stellar $V\sin i$ values reported in this paper, see Table~A.1 in \citetalias{kdwarfs-p1}.} Note that the granulation effects, and in fact the absolute magnitude of the inhomogeneities, are argued to increase with effective temperature (Sect.~\ref{s:3dmodel}) and also with decreasing metallicity, although probably the latter effect is significantly smaller \citep[e.g.,][]{allende99}, while low projected rotational velocities ($V\sin i<1$~km~s$^{-1}$) were shown not to affect the shape of the line bisectors significantly (Fig.~\ref{f:testrot3d}).

In this context, it is interesting to see the case of $\epsilon$~Eri ($\teff\simeq5050$\,K), which is more than 200~K warmer than the 3D model and yet does not show bisectors with a span significantly larger than the 3D model predictions, in particular for the strong line case. Part of the reason for this is the high $V\sin i$ value of the star (approximately 2~km~s$^{-1}$), which, as seen in Fig.~\ref{f:testrot3d}, can reduce somewhat the span of the line bisectors. Although it is not shown in Fig.~\ref{f:bis_theo_obs}, a similar result is found for the stars HIP~26779, HIP~88601, HIP~64797, and HIP~37349, which are also hotter than the 3D model by 50 to 300\,K and have high $V\sin i$ values (between 2 and 3~km~s$^{-1}$). Interestingly, the 5 stars mentioned above are also the most active in our sample (cf.~Fig.~B.1 in \citetalias{kdwarfs-p1}), which suggests that chromospheric activity or some other activity-related effect has an important impact on the shapes of \fei\ line profiles.

HIP~86400 ($\teff\simeq4830$\,K, $\feh=-0.05$) is our reference star because it has parameters very close to those adopted in the calculation of the 3D model atmosphere. A visual inspection of the third panel in Fig.~\ref{f:bis_theo_obs} shows that the agreement between the observed and predicted mean line bisectors is excellent. The line bisector spans for the strongest \fei\ features in our HIP~86400 spectrum are about 70~\ms. Since this value agrees very well with the theoretical prediction, after accurately taking into account the effects of finite spectral resolution, detailed instrumental profile, and projected rotational velocity, we can conclude that the \textit{actual} bisector spans in this type of star (solar-metallicity early K-dwarfs) are around 140~\ms\ for the strongest \fei\ features. The exact \textit{observed} value will of course depend on the properties of the instrument and the value of the projected rotational velocity of the star.

The coolest stars in our sample, HIP~114622 ($\teff\simeq4740$\,K) and HIP~23311 ($\teff\simeq4640$\,K), have both relatively low $V\sin i$ ($\lesssim2$~km~s$^{-1}$) and are about 80 and 180\,K cooler than our 3D model. Their spectra are richer in spectral lines compared to those of our other sample stars and this is likely the reason why their mean bisectors show large deviations from the theoretical expectation near the continuum (the bottom panel of Fig.~\ref{f:bis_theo_obs} shows only the results for HIP~23311, which are very similar to those of HIP~114622). In fact, the mean bisectors approach zero velocity there, which points to the influence of numerous blends randomly distributed on both sides of the spectral lines. Nonetheless, farther from the continuum, the 3D model predictions agree quite well with the observations of these two cooler stars. Therefore, it is likely that the granulation contrast and convective velocities do not change significantly in dwarf stars of effective temperature between 4600 and 4800~K.

A comment should be added here about the theoretical mean bisector of the strongest lines. Since their core wavelengths are shifted by up to +100~\ms, and we concluded this to be a limitation of our model (Sect.~\ref{s:bisectorswavshiftsnobro}), the lower part of their line bisectors should be similarly affected. The strongest features that were used to determine the mean line bisectors shown in Fig.~\ref{f:bis_theo_obs} have $EW\simeq140$\,m\AA, which is higher than the upper limit for the reliability of our 3D model results ($EW\simeq100$\,m\AA). The bisectors of $EW\simeq140$\,m\AA\ \fei\ lines are well represented in the second panel of Fig.~\ref{f:meanbist}. Note that the low EP lines are the most affected by the incorrect redshifts predicted by our 3D model whereas the high EP lines have, on average, zero shift. When plotted with respect to their core wavelengths (i.e., shifted so that the lowest flux pixel is at zero velocity), low and high EP lines show somewhat similar bisectors in this group of lines. Interestingly, of all observed lines used to construct Fig.~\ref{f:bis_theo_obs}, about 80\,\% have high EP ($>3.5$\,eV). Thus, 
this comparison between theory and observation is mostly unaffected by the model uncertainties because, for this group of lines, both high and low EP lines show similar \textit{relative} line bisectors (therefore the theoretical average strong line bisector shown in Fig.~\ref{f:bis_theo_obs} does not change significantly if computed separately for each EP group) and the observations are dominated by high EP lines, which, when computed with our 3D model, are less affected, in an absolute sense, by these limitations.

\subsection{Wavelength shifts}

The line core wavelength shift vs.~equivalent width relations for four of our sample stars are shown in Fig.~\ref{f:lshmk_theo}, along with the model predictions. In this figure, we have forced the average observed velocity for the strongest lines ($EW\gtrsim100$\,m\AA) to be zero (cf.~\citetalias{kdwarfs-p1}). The basic signature of granulation is evident in all cases and it is reasonably consistent between theory and observation, as long as we consider lines of $EW<100$~m\AA, given that the theoretical predictions for the line shifts are uncertain at large $EW$ (Sect.~\ref{s:bisectorswavshiftsnobro}).

For HIP~96100, the hottest star of the sample, a visual inspection of Fig.~\ref{f:lshmk_theo} shows that the blueshifts are underestimated by the model below $EW\simeq100$~m\AA. This is most likely due to the higher effective temperature of the star compared to that of the model. In general, we expect the slope in this relation to increase with increasing temperature.

The observed line shift vs.~equivalent width relation for $\epsilon$~Eri is more shallow than that of the 3D model and perhaps even more shallow also than those of the cooler stars shown in Fig.~\ref{f:lshmk_theo}. In addition, there is significant scatter, even though this is the brightest star of the sample and we expect random errors to be minimum (in addition to the larger scatter of the bin-averaged values shown in Fig.~\ref{f:lshmk_theo}, the error bars that represent the line-by-line scatter are also larger compared to those of fainter stars). These properties are also seen in other high $V\sin i$ stars with clear signatures of strong chromospheric activity in our sample (these stars are not shown in Fig.~\ref{f:lshmk_theo} but in this context they behave similarly to $\epsilon$~Eri). Interestingly, stellar activity has been shown to affect the shapes of the line cores significantly \citep[e.g.,][]{gray84:zeeman,borrero08}, which explains our results, at least qualitatively.

For HIP~86400, the reference star, the agreement between model and observation is excellent for lines of $EW<100$~m\AA. The predicted slope of the relation for weak lines is remarkably similar to that given by the observational data. The maximum observed convective blueshift is about $-150$\,\ms. Note that the theoretical relation does not flatten completely below 100\,m\AA\ but it seems to become slightly less dependent on $EW$ for the weakest lines $EW<40$\,m\AA. This is also in very good agreement with the observations, even though blends introduce significant error to our measurements for the weakest lines.

The agreement is also good for the cooler stars HIP~114622 and HIP~23311 (not shown), confirming our suspicion that the characteristics of granulation are quite similar between these two stars and HIP~86400, despite the $\sim200$\,K difference in $\teff$.

\section{Conclusions}

A three dimensional radiative-hydrodynamical simulation has been computed for stellar parameters $\teff=4820$~K, $\log g=4.5$, and $\feh=0$, using the same prescription that has been successful at reproducing the granulation features in the Sun. This K-dwarf model also predicts a granulation pattern, i.e., a correlation between the intensity (or temperature), velocity, and density fields, which is, however, weaker than that found for hotter stars. Strong temperature contrasts at a given height occur well below the visible surface, contrary to the case of hotter stars where this happens nearer their visible surface, but the velocity field still shows significant fluctuations in high photospheric layers.

The 3D model calculations for typical K-dwarfs show a span in the theoretical line bisectors between about 10 and 250~m~s$^{-1}$, depending on line strength, while the maximum core wavelength shifts (convective blueshifts) are about $-200$~m~s$^{-1}$ for the \fei\ lines and $-600$~\ms\ for the \feii\ lines.

Line broadening due to the projected rotational velocity of stars must be taken into account when comparing the model predictions with observational data. The impact of the finite resolving power of our observations (presented in the first part of this series) was carefully analyzed and taken into account before comparing theory to observations. We explored the effect of both a variable FWHM and an asymmetric instrumental profile. The former has only a small impact on the line bisectors and wavelength shifts (within the range of spectral resolution of this work) but the latter produce significant additional shifts to the core wavelengths of the lines.

Although there is a possibility that other effects, for example those related to stellar activity, modify the detailed shapes of spectral lines profiles, the agreement between the 3D model predictions and the observations is satisfactory. If there are effects other than granulation affecting the line profiles, we expect their impact to be smaller.

This good agreement demonstrates that our 3D model for K-dwarf granulation accounts for most of the factors that determine the detailed shapes of \fei\ line profiles, as well as their strengths, as quantified by, for example, the line equivalent width. This, in turn, means that the physics invoked in these parameter-free simulations is fairly realistic. Our model is thus adequate to explore 3D effects on chemical abundance studies, which will be done in \citetalias{kdwarfs-p3}.

It should be emphasized that even though our 3D model represents significant progress relative to the 1D scenario, there is still room for improvement. For example, the simulation predicts convective redshifts of about 100~m\,s$^{-1}$ for the strongest \fei\ lines, but implications based on solar observations and their comparison to our K-dwarf data indicate that they should be nearly zero. Non-LTE effects, the chromosphere, or numerical artifacts associated with our 3D model calculations could be responsible for this discrepancy.

\begin{acknowledgements}
This work was supported in part by the Robert A. Welch Foundation of Houston, Texas. The authors acknowledge the Texas Advanced Computing Center (TACC) at The University of Texas at Austin for providing HPC resources that have contributed to the research results reported within this paper (URL: http://www.tacc.utexas.edu).
\end{acknowledgements}

\bibliographystyle{aa}

\bibliography{1741_refs}

\begin{thebibliography}{49}
\expandafter\ifx\csname natexlab\endcsname\relax\def\natexlab#1{#1}\fi

\bibitem[{{Allende~Prieto} {et~al.}(2002){Allende~Prieto}, {Asplund},
  {Garc\'{\i}a L\'opez}, \& {Lambert}}]{allende02}
{Allende~Prieto}, C., {Asplund}, M., {Garc\'{\i}a L\'opez}, R.~J., \&
  {Lambert}, D.~L. 2002, \apj, 567, 544

\bibitem[{{Allende~Prieto} {et~al.}(1999){Allende~Prieto}, {Garc{\'{\i}}a
  L{\'o}pez}, {Lambert}, \& {Gustafsson}}]{allende99}
{Allende~Prieto}, C., {Garc{\'{\i}}a L{\'o}pez}, R.~J., {Lambert}, D.~L., \&
  {Gustafsson}, B. 1999, \apj, 526, 991

\bibitem[{{Asplund}(2005)}]{asplund05:review}
{Asplund}, M. 2005, \araa, 43, 481

\bibitem[{{Asplund} {et~al.}(2005){Asplund}, {Grevesse}, \&
  {Sauval}}]{asplund05:solarabundances}
{Asplund}, M., {Grevesse}, N., \& {Sauval}, A.~J. 2005, in Astronomical Society
  of the Pacific Conference Series, Vol. 336, Cosmic Abundances as Records of
  Stellar Evolution and Nucleosynthesis, ed. T.~G. {Barnes}, III \& F.~N.
  {Bash}, 25--+

\bibitem[{{Asplund} {et~al.}(2000{\natexlab{a}}){Asplund}, {Ludwig},
  {Nordlund}, \& {Stein}}]{asplund00:resolution}
{Asplund}, M., {Ludwig}, H.-G., {Nordlund}, {\AA}., \& {Stein}, R.~F.
  2000{\natexlab{a}}, \aap, 359, 669

\bibitem[{{Asplund} {et~al.}(2000{\natexlab{b}}){Asplund}, {Nordlund},
  {Trampedach}, {Allende~Prieto}, \& {Stein}}]{asplund00:iron_shapes}
{Asplund}, M., {Nordlund}, {\AA}., {Trampedach}, R., {Allende~Prieto}, C., \&
  {Stein}, R.~F. 2000{\natexlab{b}}, \aap, 359, 729

\bibitem[{{Asplund} {et~al.}(1999){Asplund}, {Nordlund}, {Trampedach}, \&
  {Stein}}]{asplund99}
{Asplund}, M., {Nordlund}, {\AA}., {Trampedach}, R., \& {Stein}, R.~F. 1999,
  \aap, 346, L17

\bibitem[{{Barklem} {et~al.}(2000){Barklem}, {Piskunov}, \&
  {O'Mara}}]{barklem00}
{Barklem}, P.~S., {Piskunov}, N., \& {O'Mara}, B.~J. 2000, \aaps, 142, 467

\bibitem[{{Blackwell} {et~al.}(1976){Blackwell}, {Ibbetson}, {Petford}, \&
  {Willis}}]{blackwell76}
{Blackwell}, D.~E., {Ibbetson}, P.~A., {Petford}, A.~D., \& {Willis}, R.~B.
  1976, \mnras, 177, 219

\bibitem[{{Borrero}(2008)}]{borrero08}
{Borrero}, J.~M. 2008, \apj, 673, 470

\bibitem[{{Bray} {et~al.}(1984){Bray}, {Loughhead}, \& {Durrant}}]{bray84}
{Bray}, R.~J., {Loughhead}, R.~E., \& {Durrant}, C.~J. 1984, {The solar
  granulation (2nd edition)} (Cambridge and New York, Cambridge University
  Press, 1984)

\bibitem[{{Chiavassa}(2008)}]{chiavassa08}
{Chiavassa}, A. 2008, in EAS Publications Series, Vol.~28, EAS Publications
  Series, ed. S.~{Wolf}, F.~{Allard}, \& P.~{Stee}, 31--36

\bibitem[{{Collet} {et~al.}(2006){Collet}, {Asplund}, \&
  {Trampedach}}]{collet06}
{Collet}, R., {Asplund}, M., \& {Trampedach}, R. 2006, \apjl, 644, L121

\bibitem[{{Collet} {et~al.}(2007){Collet}, {Asplund}, \&
  {Trampedach}}]{collet07}
{Collet}, R., {Asplund}, M., \& {Trampedach}, R. 2007, \aap, 469, 687

\bibitem[{{Dravins}(1987{\natexlab{a}})}]{dravins87:line_asymmetries}
{Dravins}, D. 1987{\natexlab{a}}, \aap, 172, 211

\bibitem[{{Dravins}(1987{\natexlab{b}})}]{dravins87:observability}
{Dravins}, D. 1987{\natexlab{b}}, \aap, 172, 200

\bibitem[{{Dravins}(2008)}]{dravins08}
{Dravins}, D. 2008, \aap, 492, 199

\bibitem[{{Dravins} {et~al.}(1981){Dravins}, {Lindegren}, \&
  {Nordlund}}]{dravins81}
{Dravins}, D., {Lindegren}, L., \& {Nordlund}, {\AA}. 1981, \aap, 96, 345

\bibitem[{{Freytag} {et~al.}(1996){Freytag}, {Ludwig}, \&
  {Steffen}}]{freytag96}
{Freytag}, B., {Ludwig}, H.-G., \& {Steffen}, M. 1996, \aap, 313, 497

\bibitem[{{Gilliland} \& {Dupree}(1996)}]{gilliland96}
{Gilliland}, R.~L. \& {Dupree}, A.~K. 1996, \apjl, 463, L29

\bibitem[{{Gray}(1982)}]{gray82}
{Gray}, D.~F. 1982, \apj, 255, 200

\bibitem[{{Gray}(1984)}]{gray84:zeeman}
{Gray}, D.~F. 1984, \apj, 277, 640

\bibitem[{{Gray}(1986)}]{gray86}
{Gray}, D.~F. 1986, \pasp, 98, 319

\bibitem[{{Gray}(2005)}]{gray05}
{Gray}, D.~F. 2005, \pasp, 117, 711

\bibitem[{{Gray} {et~al.}(2008){Gray}, {Carney}, \& {Yong}}]{gray08}
{Gray}, D.~F., {Carney}, B.~W., \& {Yong}, D. 2008, \aj, 135, 2033

\bibitem[{{Gray} \& {Nagel}(1989)}]{gray89}
{Gray}, D.~F. \& {Nagel}, T. 1989, \apj, 341, 421

\bibitem[{{Gray} \& {Toner}(1985)}]{gray85}
{Gray}, D.~F. \& {Toner}, C.~G. 1985, \pasp, 97, 543

\bibitem[{{Grevesse} \& {Sauval}(1998)}]{grevesse98}
{Grevesse}, N. \& {Sauval}, A.~J. 1998, Space Science Reviews, 85, 161

\bibitem[{{Gustafsson} {et~al.}(1975){Gustafsson}, {Bell}, {Eriksson}, \&
  {Nordlund}}]{gustafsson75}
{Gustafsson}, B., {Bell}, R.~A., {Eriksson}, K., \& {Nordlund}, {\AA}. 1975,
  \aap, 42, 407

\bibitem[{{Gustafsson} {et~al.}(2008){Gustafsson}, {Edvardsson}, {Eriksson},
  {J{\o}rgensen}, {Nordlund}, \& {Plez}}]{gustafsson08}
{Gustafsson}, B., {Edvardsson}, B., {Eriksson}, K., {et~al.} 2008, \aap, 486,
  951

\bibitem[{{Hubeny}(1988)}]{hubeny88}
{Hubeny}, I. 1988, Computer Phys. Comm., 52, 103

\bibitem[{{Hubeny} \& {Lanz}(1995)}]{hubeny95}
{Hubeny}, I. \& {Lanz}, T. 1995, \apj, 439, 875

\bibitem[{{Janssen} \& {Cauzzi}(2006)}]{janssen06}
{Janssen}, K. \& {Cauzzi}, G. 2006, \aap, 450, 365

\bibitem[{{Koesterke} {et~al.}(2008){Koesterke}, {Allende~Prieto}, \&
  {Lambert}}]{koesterke08}
{Koesterke}, L., {Allende~Prieto}, C., \& {Lambert}, D.~L. 2008, \apj, 680, 764

\bibitem[{{Kurucz}(1979)}]{kurucz79}
{Kurucz}, R.~L. 1979, \apjs, 40, 1

\bibitem[{{Kurucz}(1993{\natexlab{a}})}]{kurucz93:cd13}
{Kurucz}, R.~L. 1993{\natexlab{a}}, ATLAS9 Stellar Atmosphere Programs and 2
  km/s grid.~Kurucz CD-ROM No.~13.~ Cambridge, Mass.: Smithsonian Astrophysical
  Observatory, 1993., 13

\bibitem[{{Kurucz}(1993{\natexlab{b}})}]{kurucz93:cd18}
{Kurucz}, R.~L. 1993{\natexlab{b}}, SYNTHE Spectrum Synthesis Programs and Line
  Data.~Kurucz CD-ROM No.~18.~Cambridge, Mass.: Smithsonian Astrophysical
  Observatory, 1993., 18

\bibitem[{{Ludwig} {et~al.}(1999){Ludwig}, {Freytag}, \& {Steffen}}]{ludwig99}
{Ludwig}, H.-G., {Freytag}, B., \& {Steffen}, M. 1999, \aap, 346, 111

\bibitem[{{Mihalas} {et~al.}(1988){Mihalas}, {Dappen}, \& {Hummer}}]{mihalas88}
{Mihalas}, D., {Dappen}, W., \& {Hummer}, D.~G. 1988, \apj, 331, 815

\bibitem[{{Muller}(1999)}]{muller99}
{Muller}, R. 1999, in Astrophysics and Space Science Library, Vol. 239, Motions
  in the Solar Atmosphere, ed. A.~{Hanslmeier} \& M.~{Messerotti}, 35--70

\bibitem[{{Nordlund} \& {Dravins}(1990)}]{nordlund90}
{Nordlund}, {\AA}. \& {Dravins}, D. 1990, \aap, 228, 155

\bibitem[{{Nordlund} {et~al.}(2008){Nordlund}, {Stein}, \&
  {Asplund}}]{nordlund08}
{Nordlund}, A., {Stein}, R.~F., \& {Asplund}, M. 2008, {L}iving Reviews in
  Solar Physics, in press

\bibitem[{{Ram\'irez} {et~al.}(2009){Ram\'irez}, {Allende~Prieto}, {Koesterke},
  {Lambert}, \& {Asplund}}]{kdwarfs-p3}
{Ram\'irez}, I., {Allende~Prieto}, C., {Koesterke}, L., {Lambert}, D.~L., \&
  {Asplund}, M. 2009, in preparation (Paper~III)

\bibitem[{{Ram{\'{\i}}rez} {et~al.}(2007){Ram{\'{\i}}rez}, {Allende~Prieto}, \&
  {Lambert}}]{ramirez07}
{Ram{\'{\i}}rez}, I., {Allende~Prieto}, C., \& {Lambert}, D.~L. 2007, \aap,
  465, 271

\bibitem[{{Ram{\'{\i}}rez} {et~al.}(2008){Ram{\'{\i}}rez}, {Allende Prieto}, \&
  {Lambert}}]{kdwarfs-p1}
{Ram{\'{\i}}rez}, I., {Allende Prieto}, C., \& {Lambert}, D.~L. 2008, \aap,
  492, 841, (Paper~I)

\bibitem[{{Robinson} {et~al.}(2003){Robinson}, {Demarque}, {Li}, {Sofia},
  {Kim}, {Chan}, \& {Guenther}}]{robinson03}
{Robinson}, F.~J., {Demarque}, P., {Li}, L.~H., {et~al.} 2003, \mnras, 340, 923

\bibitem[{{Stein} \& {Nordlund}(1998)}]{stein98}
{Stein}, R.~F. \& {Nordlund}, {\AA}. 1998, \apj, 499, 914

\bibitem[{{Trampedach}(2007)}]{trampedach07}
{Trampedach}, R. 2007, in American Institute of Physics Conference Series, Vol.
  948, American Institute of Physics Conference Series, 141--148

\bibitem[{{V{\"o}gler}(2004)}]{voegler04}
{V{\"o}gler}, A. 2004, \aap, 421, 755

\end{thebibliography}

\begin{appendix}

\section{The advanced spectrum synthesis 3D tool (\asset)} \label{s:asset}

A computer code for solving the radiative transfer problem in 3D model atmospheres has been recently developed by \cite{koesterke08} and was made available for this work.

\cite{koesterke08} ``Advanced Spectrum Synthesis 3D Tool'' (\asset) represents an improvement over previous 3D/LTE codes because it is not limited to the calculation of single lines, simple blends, or constant background opacities. Also, it properly includes electron and Rayleigh scattering due to atomic hydrogen, and uses improved (higher-order) interpolation schemes within the simulation grid points for the calculation of opacities and intensities. \asset\ is also capable of working with 1D models, which allows 1D vs.~3D comparisons to be made in a consistent manner. Furthermore, the opacity calculation is based on a modified version of the thoroughly tested 1D code {\tt SYNSPEC} \citep{hubeny88,hubeny95}.

\begin{figure}
\includegraphics[bb=85 360 620 695,width=9.3cm]{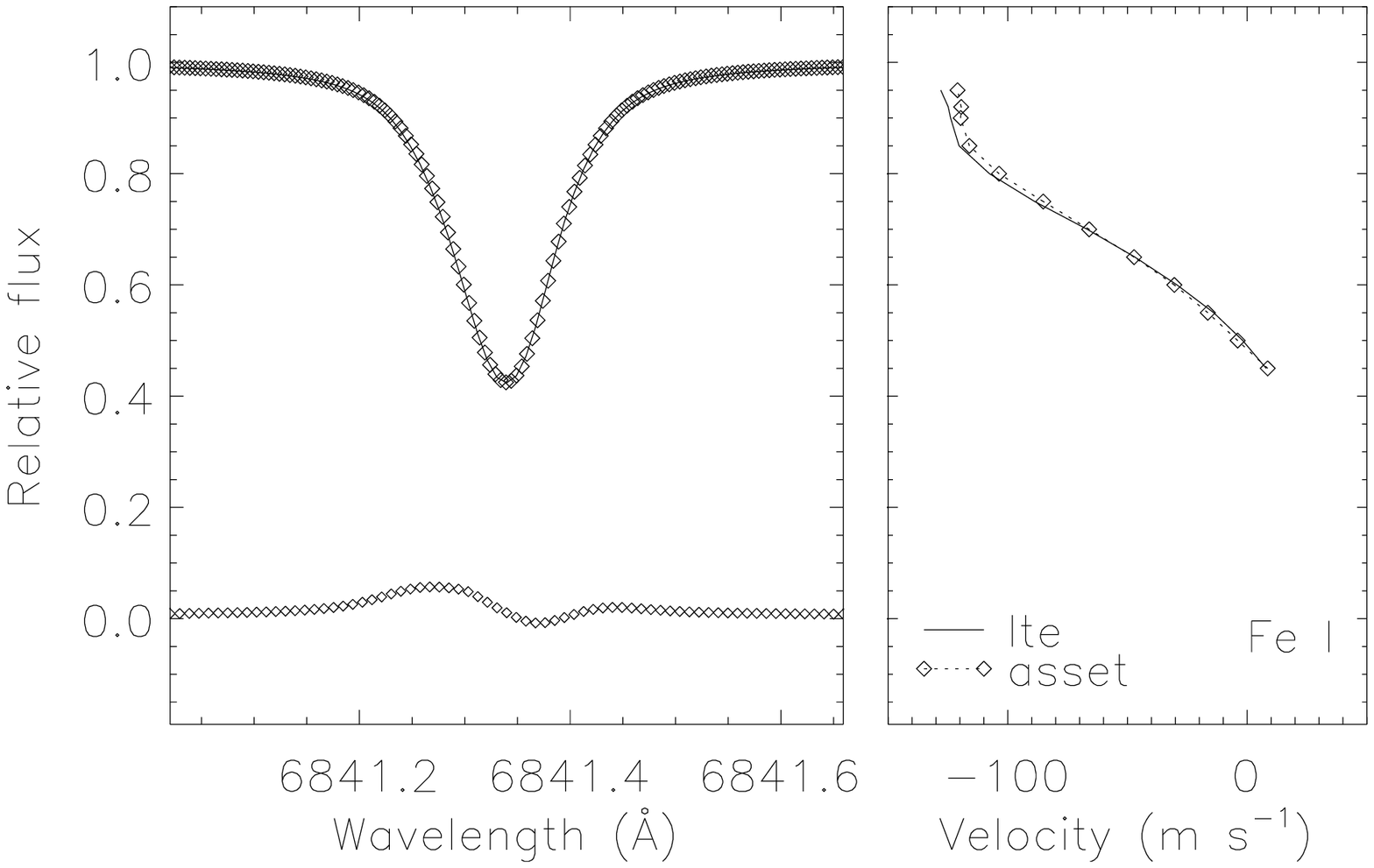}
\caption{Left: Spectral lines synthesized with the 3D codes {\tt lte} (solid line) and \asset\ (diamonds). Their difference ({\tt lte} minus \asset) is shown at the bottom, amplified by a factor of 10. Right: Bisectors of the lines shown on the left panel.}
\label{f:assetvslte_line}
\end{figure}

\begin{figure}
\includegraphics[bb=54 370 350 865,width=9cm]{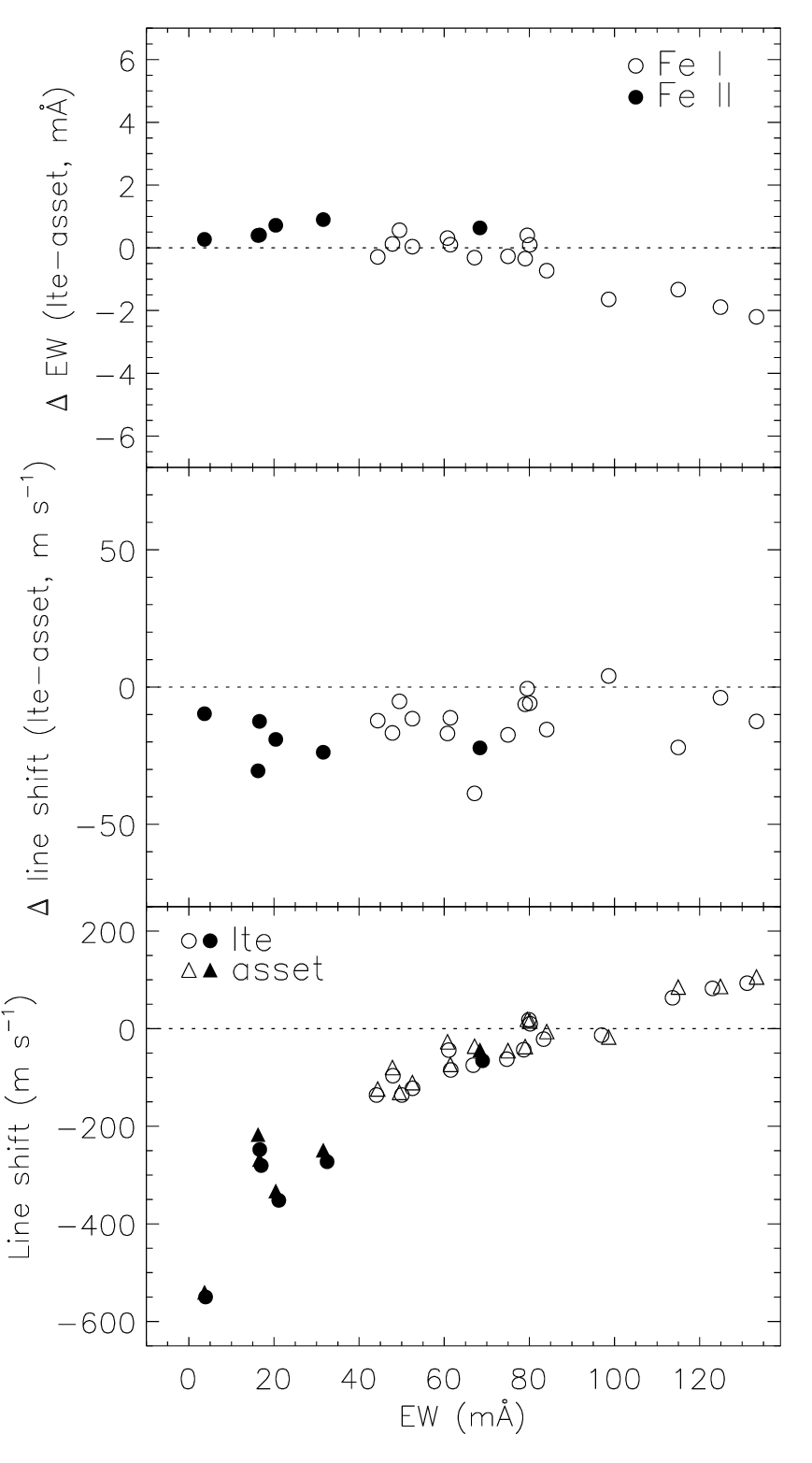}
\caption{Difference in equivalent width (top panel), line shift (middle panel), and EW vs.~line shift relation (bottom panel) between the predictions of {\tt lte} and \asset. Filled (open) symbols correspond to \feii\ (\fei) lines. In the bottom panel, the triangles (circles) correspond to the \asset\ ({\tt lte}) predictions.}
\label{f:assetvslte_ewlsh}
\end{figure}

In Paper III, we will use \asset\ to calculate molecular and atomic features as well as the continuum over wide spectral windows. For the synthesis of atomic lines in this paper, however, we used the 3D code ``{\tt lte}'' \citep[e.g.,][]{asplund00:iron_shapes}, as described in Sect.~\ref{s:3dsynthesis}. We computed 23 of the spectral lines used in this paper (17 \fei\ and 6 \feii) with \asset\ to perform a code-to-code comparison. Input data were carefully checked for consistency. A typical result of the comparison of detailed line profiles is shown in Fig.~\ref{f:assetvslte_line} while the differences in the predicted equivalent widths and core wavelength shifts are shown in Fig.~\ref{f:assetvslte_ewlsh}.

In general, the wings of line profiles computed with {\tt lte} are slightly deeper, in particular the blue wing. This makes the line bisector span slightly larger, by less than about 10\,\ms\ for the example shown in Fig.~\ref{f:assetvslte_line}. This also leads to a slightly higher (more negative) convective blueshift for the {\tt lte} case, as shown in the middle panel of Fig.~\ref{f:assetvslte_ewlsh}. The line core blueshift is higher (more negative, by about 15~\ms) according to {\tt lte}.

The top panel of Fig.~\ref{f:assetvslte_ewlsh} shows that the equivalent widths obtained from the line profiles computed with both codes agree within 1~m\AA\ below 90~m\AA\ and the difference shows a small trend for the strongest lines, with the $EW$ values from {\tt lte} being lower by at most 2~m\AA. These differences can be attributed not only to the slightly different detailed shapes of the profiles but also to the continuum level computed for each case.

Note that the line shift vs.~$EW$ relation is only slightly affected by the choice of 3D spectrum synthesis code (bottom panel of Fig.~\ref{f:assetvslte_ewlsh}). Only a small shift upwards of 15~\ms\ should be considered if \asset\ is used instead of {\tt lte}. Since both the line bisectors and line shift vs.~$EW$ relation are nearly independent on the synthesis code adopted, our conclusions about the good agreement between theory and observation presented in Sect.~\ref{s:comparison} are robust.

\section{The projected rotational velocity of HIP~86400} \label{s:vsini86}

By convolving line profiles predicted by our 3D model with projected rotational velocity and instrumental profiles, as described in Sect.~\ref{s:davprof}, we determined a more accurate $V\sin i$ value for HIP~86400. We did not perform similar calculations for the other sample stars because the strength and FWHM of the lines are very sensitive to the stellar parameters, in particular $\teff$, and only HIP~86400 has parameters identical (within observational errors) to those of our 3D model atmosphere.

\begin{figure}
\includegraphics[bb=80 375 560 700,width=8.9cm]{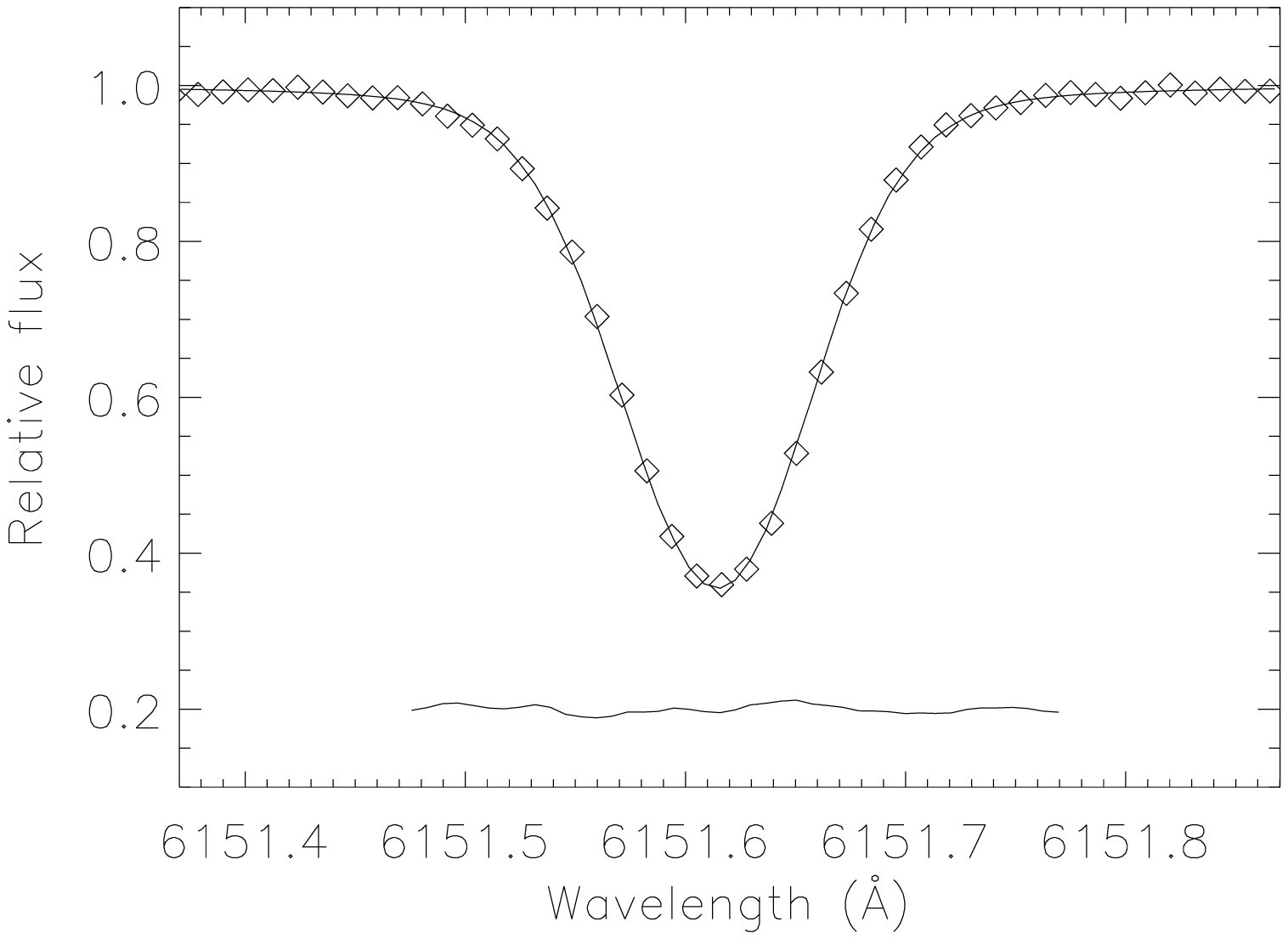}
\caption{Minimum $\chi^2$ model fit (solid line) to the observed 6151~\AA\ line profile (diamonds) in our HIP~86400 spectrum. Residuals around $\pm0.15$~\AA\ of the line center are shown at 0.2 relative flux.}
\label{f:vsini86}
\end{figure}

We selected 14 of the ``cleanest'' iron lines available in the HIP~86400 spectrum and computed the difference between observed and predicted profiles around $\pm0.15$~\AA\ from the line center using only two free parameters, the $\log gf$ value (or, equivalently, iron abundance) and $V\sin i$. By minimizing the difference between observed and predicted profiles, using a $\chi^2$-like scheme, we determined the $V\sin i$ value of HIP~86400. No additional broadening, in particular microturbulence or macroturbulence, was necessary to accurately fit the observed line profiles. One of our best fits to the data is illustrated in Fig.~\ref{f:vsini86}. The average of the $V\sin i$ values obtained from all lines, weighted by the quality of each fit, is $V\sin i=1.57\pm0.20$~km~s$^{-1}$ (the error bar corresponds to the 1-$\sigma$ scatter of the line-by-line values).

\end{appendix}

\end{document}